\documentstyle[12pt]{article}
\begin{document} 

\begin{titlepage}

\hrule 
\leftline{}
\leftline{Preprint
          \hfill   \hbox{\bf CHIBA-EP-112}}
\leftline{\hfill   \hbox{hep-th/9904045}}
\leftline{\hfill   \hbox{August 2000}}
\vskip 5pt
\hrule 
\vskip 1.0cm

\centerline{\large\bf 
A Formulation of the Yang-Mills Theory  
} 
\centerline{\large\bf  
}
\centerline{\large\bf  
as a Deformation of a Topological Field Theory 
}
\centerline{\large\bf  
}
\centerline{\large\bf  
Based on the Background Field Method 
}
\vskip 0.5cm
\centerline{\large\bf  
and Quark Confinement Problem$^*$
}
   
\vskip 1cm

\centerline{{\bf 
Kei-Ichi Kondo$^{1,}{}^{\dagger}$
}}  
\vskip 1cm
\begin{description}
\item[]{\it \centerline{ 
$^1$ Department of Physics, Faculty of Science, 
Chiba University,  Chiba 263-8522, Japan}
  }
\item[]{\centerline{$^\dagger$ 
  E-mail:  kondo@cuphd.nd.chiba-u.ac.jp }
  }
\end{description}

\centerline{{\bf Abstract}}
By making use of the background field method, we derive a novel
reformulation of the Yang-Mills theory which was proposed recently
by the author to derive quark confinement in QCD.  This
reformulation identifies the Yang-Mills theory with a deformation of
a topological quantum field theory.   The relevant background is
given by the topologically non-trivial field configuration,
especially, the topological soliton which can be identified with the
magnetic monopole current in four dimensions. We argue that the
gauge fixing term becomes dynamical and that the gluon mass
generation takes place by a spontaneous breakdown of the hidden supersymmetry
caused by the dimensional reduction.   We also propose a numerical
simulation to confirm the validity of the scheme we have proposed.
Finally we point out that the gauge fixing part may have a geometric
meaning from the viewpoint of global topology where the magnetic
monopole solution represents the critical point of a Morse function
in the space of field configurations.


\vskip 0.2cm
\hrule  

\vskip 0.5cm  

$^*$ To be published in Intern. J. Mod. Phys. A.

\end{titlepage}

\pagenumbering{arabic}

\section{Introduction}
\setcounter{equation}{0}
\par

More than twenty years ago many authors \cite{DSC} have proposed
various strategies of deriving quark confinement in quantum
chromodynamics (QCD).  One of them is to show that the QCD vacuum is
the dual superconductor which squeezes the color electric flux
between quarks and anti-quarks.  The evidences have been accumulated
by recent investigations.  Especially, recent numerical simulations
have confirmed this picture, see \cite{review,Bali98}.   In this
scenario, the magnetic monopole
\cite{Dirac31,WY75,GO78} obtained by the Abelian projection
\cite{tHooft81} in QCD plays the essential role
\cite{EI82,SY90}. These results suggest that the low-energy
effective theory of QCD is given by the dual Ginzburg-Landau theory
\cite{Suzuki88}. In fact, it has been shown that the dual
Ginzburg-Landau theory can be derived starting from the QCD
Lagrangian at least in the strong coupling region, see e.g,
\cite{KondoI}.  
\par
In the previous paper \cite{KondoII}, we have proposed a novel
formulation of the Yang-Mills theory as a
(perturbative) deformation of a topological quantum field theory
(TQFT).%
\footnote{In this reformulation, the gauge-fixed Yang-Mills theory is decomposed into the TQFT part and the remaining part.  Then we assume that the remaining part can be treated in perturbation theory in the gauge coupling constant.  This assumption is nothing but the meaning of the perturbative deformation. 
}    
We have shown \cite{KondoII,KondoIV} that the quark confinement in
QCD in the sense of area law of the Wilson loop (or
equivalently, the linear static potential between quark and
anti-quark) can be derived from the formulation at least in the
maximal Abelian (MA) gauge.   The MA gauge realizes the Abelian
magnetic monopole in Yang-Mills theory without introducing the
scalar field as an elementary field (Hence it is realized as a
composite field constructed from the gauge degrees of freedom).  In
the similar way, it has been shown
\cite{KondoIII} that the four-dimensional Abelian gauge theory can
have the confining phase in the strong coupling region. 
This can be used to give another derivation of quark confinement in
QCD based on the low-energy effective {\it Abelian} gauge theory,
see \cite{KondoV}.
\par
The above results are consistent with those of lattice gauge
theory
\cite{LGT}, though our formulation is given directly on the
continuum space-time.   This similarity is due to a fact that the
ingredients of confinement in our formulation lies in the
compactness of the gauge group (or the periodicity in the gauge
potential) and the existence of topological soliton. Thus, the
existence of magnetic monopole is a sufficient condition for
explaining  quark confinement, as confirmed by analytical and
numerical results
\cite{review}. 
\par
In this paper, we re-derive the formulation proposed in
\cite{KondoII,KondoIII} based on the background field method (BGFM)
\cite{DeWitt67,tHooft75,Abbott82,Abbott81,AGS83}.
A purpose of this paper is to fill the gap in the previous
presentation
\cite{KondoII} without any ad hoc argument.  This derivation enables
us to discuss various topological soliton or topological defect
other than the magnetic monopole, which might equally play the
important role in explaining  the origin of quark confinement.  
Such a viewpoint is necessary to answer the question: what are the
most relevant degrees of freedom for quark confinement, since the
necessary and sufficient condition for quark confinement is not yet
known.  Therefore, our formulation can also be applied to other
scenarios of quark confinement based on various confiners, e.g.
instanton, center vortex or non-Abelian magnetic monopole, although
the details will be given in a subsequent paper.
Another advantage of BGFM is that it simplifies the proof
\cite{QR97,KondoI} that the renormalization group beta function of
the Abelian-projected effective gauge theory is the same as the
original Yang-Mills non-Abelian gauge theory.
\par
Just as the topological Yang-Mills theory
\cite{Witten88} describes the gauge field configurations satisfying
the self-dual equation, i.e., instantons,
\begin{eqnarray}
 {\cal F}_{\mu\nu}=\pm \tilde {\cal F}_{\mu\nu} ,
\quad \tilde {\cal F}_{\mu\nu} := {1 \over 2} \epsilon_{\mu\nu\rho\sigma}   {\cal F}^{\rho\sigma} ,
 \label{SDeq0}
\end{eqnarray}
the TQFT that we have proposed deals with the gauge field
configurations which obeys the MA gauge equation,  
\begin{eqnarray}
 D^{\mp}_\mu[a]A_\mu^{\pm}=0 ,
 \label{MAGeq}
\end{eqnarray}
which is nothing but the background field equation.
Both equations are the 1st order partial differential equations. 
They may have some properties in common.  In fact, a class of
classical solutions of the MA gauge equation (\ref{MAGeq})
simultaneously satisfies  the self-dual equation
(\ref{SDeq0}) and vice versa \cite{BOT96,CG95}.  It is obtained from
the same ansatz as that of 't Hooft for the multi-instanton.  The
instanton is the point defect in four dimensions, while the magnetic
monopole is the point defect in three dimensions. In four
dimensions, therefore, the magnetic monopole is a one-dimensional
object, i.e., a current $k_\mu$ (a closed loop due to the topological
conservation law $\partial_\mu k_\mu = 0$).  This is a Lorentz
covariant generalization of the observation that the static monopole
in three dimensions draws the straight line in the time direction in
four dimensions where the monopole charge is given by the integral 
$Q_m:=\int d^3x k_0(x)$ from the monopole density $k_0$. Since
we frequently use the 'magnetic' monopole as implying the solution of
the MA gauge equation, the solution can describe the object which
looks like the magnetic monopole and the instanton at the same
time.   Therefore, the magnetic monopole and the instanton are not
the disjoint concept in four dimensions.
Actually,  strong correlations between monopoles and instantons are
shown in the analytical studies \cite{BOT96,CG95} and observed in the
lattice simulations
\cite{HT96,BS96,STSM95,FSST97,FMT97}.
It is easy to see that the instanton is also a solution of the field
equation (2nd order partial differential equation)
\begin{eqnarray}
 {\cal D}_\nu[{\cal A}] {\cal F}_{\mu\nu}=0 .
 \label{Feq0}
\end{eqnarray}
However, it is not yet clarified which solution of the MA gauge
equation (\ref{MAGeq}) becomes that of the field equation besides
the solution mentioned above. The solution of (\ref{MAGeq}) may
contain the solution which is not the solution of the
field equation (\ref{Feq0}).
\par
When we see the intersection of the
magnetic monopole current with the two-dimensional plane, the
classical configuration satisfying (\ref{MAGeq})
looks like the instanton in two-dimensional nonlinear sigma model
(NLSM$_2$), as shown in
\cite{KondoII}.   Therefore the condensation of the magnetic monopole
current in four dimensions can be examined on the two-dimensional
subspace which can be chosen arbitrarily. The condensation of the
two-dimensional instanton in NLSM$_2$ leads to that of the
four-dimensional magnetic monopole current in Yang-Mills theory.  So
the instanton condensation in NLSM$_2$ is a sufficient condition of
quark confinement based on the dual superconductor scenario.
\par
In the scenario \cite{KondoII} of deriving quark confinement,
the gauge fixing part for the gauge fixing condition (\ref{MAGeq})
has played the essential role.  
Since the quark confinement must be a gauge invariant concept, it
is better to derive it based on the gauge invariant formulation.  In
contrast to the lattice gauge theory, however, the continuum
formulation of the ordinary gauge theory free from the gauge fixing
is not available except for special cases.   Then we are forced to
deal with the formulation based on the specific choice of gauge
fixing.  The readers might think the claim strange that the essence
of quark confinement lies in the gauge-fixing part. 
Recall that, in the level of the classical theory, the action part of
the gauge theory is well understood from a viewpoint of the geometry
of connection.   In quantum theory, however, we need to include the
gauge fixing term in order to correctly quantize the gauge theory. 
Usually, the gauge fixing term introduced in this way is not
considered to have any geometric meaning.  
However, this observation is
not necessarily correct.  In fact, the gauge fixing term plus the
associated Faddeev-Popov ghost term can have the very
geometric meaning from the viewpoint of global topology, as will
be discussed in this paper.  In the quantum gauge theory,
therefore, the action part and the gauge fixing part should be
treated on equal footing. Unfortunately, we must discuss the
topology  of the infinite dimensional manifold for gauge field
configurations.  Then the mathematically rigorous analysis will be
rather hard, so that we can at best analyze the finite dimensional
analog.
\par
Usually, we consider that, even if the gauge fixing term has a
geometric interpretation, it can not have any local dynamics
(propagating mode) and describe only the topological objects, since
it is written as the Becchi-Rouet-Stora-Tyupin (BRST) exact form,
i.e.,
$S_{GF}=\{ Q_B, \kappa \Psi \}$ 
using the BRST charge
$Q_B$.  In the manifestly covariant formalism of gauge theory, the
physical state
$| phys \rangle$ is specified by the condition,
$Q_B | phys \rangle =0$.
If we consider the theory with the action 
$S_{GF}=\{ Q_B, \kappa \Psi \}$ alone by neglecting the Yang-Mills
action (this theory is  identified with the TQFT), the
expectation value of the gauge invariant quantity 
$\langle {\cal O} \rangle$
does not depend on the coupling $\kappa$, since
${\partial \over \partial \kappa} \langle 0|{\cal O} |0 \rangle
= - \langle 0| {\cal O} \{ Q_B, \Psi \} |0 \rangle 
+ \langle 0| {\cal O} |0 \rangle \langle 0|\{ Q_B, \Psi \} |0 \rangle
= - \langle  0| \{ Q_B, {\cal O} \Psi \} \rangle 
= 0
$
where we have used the BRST invariance of ${\cal O}$ and 
$|0 \rangle \in |phys \rangle$.
However, taking into account the action $S_{YM}$ in addition to
$S_{GF}$, we can not draw the same conclusion.  This is the usual
situation of quantized gauge theory.
A subtle point is that the above consideration is based on the
assumption that the BRST symmetry is not broken.
If the BRST symmetry happen to be spontaneously broken
\cite{Fujikawa83}, the physical state including the vacuum is not
annihilated by the BRST charge,  i.e., $Q_B | phys \rangle \not=0$. 
In this case, the TQFT with the action 
$S_{TQFT}=\{ Q_B, \kappa \Psi \}$ can have local dynamics and the
expectation value can depend on the coupling constant $\kappa$. 
\par
In our scenarios, the spontaneous breaking of the hidden supersymmetry $OSp(4|2)$ rather than the BRST symmetry
can take place by the dimensional reduction (in the sense of
Parisi-Sourlas \cite{PS79,HK85}), at least for a special choice of
the MA gauge
\cite{KondoII}.   Here it is worth remarking that the equivalence of the correlation
functions hold only for a class of them and hence the Hilbert space
of the reduced theory is different from the original theory
\cite{KondoII}.   The symmetry breaking occurs spontaneously in the following
sense. We can choose arbitrary
$(D-2)$-dimensional subspace from the $D$-dimensional spacetime.  Once
the specific subspace is chosen, however, the hidden supersymmetry $OSp(4|2)$, i.e., the rotational symmetry in the superspace  is
broken by this procedure.

\par
Another purpose of this paper is to propose a numerical simulation
in order to confirm the dimensional reduction and
examine its implications to quark confinement problem.  The result
will prove or disprove the validity of our scenario for deriving
quark confinement based on the above reformulation.

\par
This paper is organized as follows.  
In section 2, we briefly review the BGFM for the
Yang-Mills theory and its BRST version based on the functional
integral formalism. 
In section 3, we explain how the quantum theory of
topological soliton can be obtained in the framework of BGFM.
We discuss a relationship between the instanton and the magnetic
monopole in this construction.
In section 4, by making the change of gauge field variable, we
show that the formulation proposed in \cite{KondoII} is recovered
from the BFGM.  This is the main result of this paper.
In section 5, we give a strategy of deriving quark confinement based
on the above formulation.  We take up some issues which have not
been mentioned in the previous publications.  We give a
proposal of numerical calculation for checking the validity of the
strategy. 
In section 6, we examine the mass generation for the gluon field in
the MA gauge.  We discuss a possibility of mass generation caused
by the dimensional reduction as a result of breakdown of the
hidden supersymmetry. In section 7, we discuss that the gauge fixing part in
the quantum theory of gauge fields can have a geometric meaning
from the viewpoint of global topology. In the final section, we
summarize the results and discuss the role of various topological
solitons other than the magnetic monopole for explaining color
confinement in QCD.

\section{Background field method}
\setcounter{equation}{0}
\par

\subsection{Path integral for Yang-Mills field}

We consider the functional integral approach to the Yang-Mills 
gauge field theory with the action
\begin{eqnarray}
 S_{YM}[{\cal A}] := \int d^Dx {\cal L}_{YM}[{\cal A}] 
 = - \int d^Dx {1 \over 4} ({\cal F}_{\mu\nu}^A[{\cal A}])^2 ,
\end{eqnarray}
where ${\cal F}_{\mu\nu}^A[{\cal A}]$ is the field strength for the
gauge field ${\cal A}_\mu^A$ defined by
\begin{eqnarray}
 {\cal F}_{\mu\nu}^A[{\cal A}] 
 := \partial_\mu {\cal A}_\nu^A - \partial_\nu {\cal A}_\mu^A
 + g f^{ABC} {\cal A}_\mu^B {\cal A}_\nu^C .
\end{eqnarray}
\par
In the quantum theory of the Yang-Mills gauge field, the generating
functional is defined by
\footnote{The tilde is used only for later convenience (in section
4) and does not have particular physical meaning.
}
\begin{eqnarray}
 Z[J] := \int [d{\cal A}] \delta(\tilde F^A[{\cal A}]) 
 \det \left[ {\delta \tilde F^A \over \delta \tilde \omega^B} \right]
 \exp \left\{ i [ S_{YM}[{\cal A}] + (J_\mu \cdot {\cal A}_\mu) ]
\right\} ,
\label{Z[J]}
\end{eqnarray}
where $(J \cdot {\cal A})$ is the source term
\begin{eqnarray}
  (J_\mu \cdot {\cal A}_\mu):= \int d^Dx J_\mu^A(x) {\cal A}_\mu^A(x)
.
\end{eqnarray}
In (\ref{Z[J]}), the gauge-fixing condition is imposed by
\begin{eqnarray}
  \tilde F^A[{\cal A}] = 0 ,
\end{eqnarray}
and 
$
\det \left[ {\delta \tilde F^A \over \delta \tilde \omega^B} \right]
$ is the so-called Faddeev-Popov (FP) determinant
which is the determinant of the derivative of the gauge-fixing
function $\tilde F^A$ under an infinitesimal gauge transformation,
\begin{eqnarray}
 \delta {\cal A}_\mu^A
 =  {\cal D}_\mu^{AB}[{\cal A}] \tilde \omega^B
 :=    \partial_\mu \tilde \omega^A 
 + gf^{ABC} {\cal A}_\mu^B \tilde \omega^C  ,
 \\
 {\cal D}_\mu^{AB} := \partial_\mu \delta^{AB} 
 - gf^{ABC} {\cal A}_\mu^C .
\end{eqnarray}
The delta function is made less singular by introducing the
gauge-fixing parameter $\tilde \alpha$ as
\begin{eqnarray}
 \delta(\tilde F^A[{\cal A}]) 
 := \prod_{x, A} \delta(\tilde F^A[{\cal A}(x)])
 \rightarrow \exp \left\{ -i{1 \over 2\tilde \alpha} 
 (\tilde F[{\cal A}] \cdot \tilde F[{\cal A}])
\right\} .
\end{eqnarray}
\par
For example, a common choice is the Lorentz gauge,
\begin{eqnarray}
  \tilde F^A[{\cal A}] = \partial_\mu {\cal A}_\mu^A .
\end{eqnarray}
Then the FP determinant is given by
\begin{eqnarray}
\det \left[ {\delta \tilde F^A \over \delta \tilde \omega^B} \right]
= \det (\partial_\mu {\cal D}_\mu^{AB}[{\cal A}]\delta^D(x-y)) .
\end{eqnarray}
The connected Green's functions are generated by
\begin{eqnarray}
 W[J] := - i \ln Z[J] .
\end{eqnarray}
The effective action is defined by making the Legendre transformation
\begin{eqnarray}
 \Gamma[\bar Q] := W[J] - (J_\mu \cdot \bar Q_\mu) ,
\end{eqnarray}
where
\begin{eqnarray}
 \bar Q^A := {\delta W \over \delta J_\mu^A} .
\end{eqnarray}
It is well known that the derivative of the effective action with
respect to
$\bar Q$ are the one-particle irreducible (1PI) Green's function.

\subsection{BGFM}
\par
Next, we consider the quantization on a given background gauge field
$\Omega_\mu$,
\begin{eqnarray}
 {\cal A}_\mu = \Omega_\mu + {\cal Q}_\mu ,
 \label{separa}
\end{eqnarray}
where ${\cal Q}_\mu$ denotes the field to be quantized.
The generating functional is given by
\begin{eqnarray}
 \tilde Z[J, \Omega] 
 := \int [d{\cal Q}] 
 \det \left[ {\delta \tilde F^A \over \delta \tilde \omega^B} \right]
 \exp \left\{ i \left[ S_{YM}[\Omega+{\cal Q}] 
 + (J_\mu \cdot {\cal Q}_\mu) 
 -{1 \over 2\tilde \alpha} 
 (\tilde F[{\cal Q}] \cdot \tilde F[{\cal Q}]) \right]
\right\} .
\label{Z[J,O]}
\end{eqnarray}
where the gauge invariance for
${\cal Q}_\mu$ is broken by the the gauge fixing condition 
$\tilde F^A[{\cal Q}]=0$ which is
supposed to fix completely the gauge degrees of freedom and 
the ${\delta \tilde F^A \over \delta \tilde \omega^B}$ is the
derivative of the gauge-fixing term under the infinitesimal gauge
transformation given by
\begin{eqnarray}
  \delta {\cal Q}_\mu^A 
  =   ({\cal D}_\mu[\Omega+{\cal Q}] \tilde \omega)^A  .
\end{eqnarray}
In (\ref{Z[J,O]}), we do not couple the background field to the
source following 't Hooft \cite{tHooft75} 

In the background field method (BGFM) 
\cite{DeWitt67,tHooft75,Abbott82,Abbott81,AGS83}, the the following
gauge fixing condition is chosen,
\begin{eqnarray}
  \tilde F^A[{\cal Q}] := {\cal D}_\mu^{AB}[\Omega] {\cal Q}_\mu^B 
  = 0  ,
  \label{BGFgauge}
\end{eqnarray}
which is called the background field (BGF) gauge.
An advantage of the BGF gauge is that the BGF gauge
condition retains explicit gauge invariance for the background
gauge field $\Omega_\mu$ even after the gauge fixing for the field
${\cal Q}_\mu$. 
\par
{\it Proposition}\cite{Abbott82}: Under the BGF gauge condition
(\ref{BGFgauge}), the BGF generating functional 
$\tilde Z[J, \Omega]$ and 
$\tilde W[J, \Omega] := -i \ln \tilde Z[J, \Omega]$
are invariant under the (infinitesimal) transformation,
\begin{eqnarray}
 \delta \Omega_\mu^A
 &=& ({\cal D}_\mu[\Omega] \omega)^A
 :=   (\partial_\mu \omega + i g [\omega,
\Omega_\mu])^A ,
 \label{traOme}
 \\
 \delta J_\mu^A &=& i g[\omega, J_\mu]^A 
 := - g f^{ABC} \omega^B J_\mu^C .
 \label{traJ}
\end{eqnarray}
\par
This is shown as follows.  By making the change of integration
variables, 
${\cal Q}_\mu \rightarrow {\cal Q}_\mu + i[\omega, {\cal
Q}_\mu]$, i.e.,
\begin{eqnarray}
  \delta {\cal Q}_\mu^A = i g[\omega, {\cal Q}_\mu]^A 
  = \omega \times {\cal Q} .
  \label{traQ}
\end{eqnarray}
Eq.~(\ref{traJ}) and (\ref{traQ})  represent an adjoint group
rotation for $J_\mu$ and ${\cal Q}_\mu$ respectively, so the term
$(J_\mu \cdot Q_\mu)$ is clearly invariant.  Adding (\ref{traOme})
and (\ref{traQ}), we find 
\begin{eqnarray}
 \delta (\Omega_\mu+{\cal Q}_\mu)^A
=   ({\cal D}_\mu[\Omega+{\cal Q}] \omega)^A .
 \label{traA}
\end{eqnarray}
This is just a gauge transformation on the field variable
${\cal A}_\mu=\Omega_\mu+{\cal Q}_\mu$, so the action
$S_{YM}[\Omega+{\cal Q}]$ is also invariant.
Note that the BGF gauge condition $\tilde F^A[{\cal Q}]$ is just
the covariant derivative of ${\cal Q}_\mu$ with respect to the BGF
$\Omega_\mu$.  Eq.~(\ref{traOme}) is a gauge transformation on
$\Omega_\mu$ and (\ref{traQ}) is an adjoint rotation of ${\cal
Q}_\mu$.  Then the gauge fixing term $(\tilde F \cdot \tilde F)$ is
invariant under such transformations.  The FP determinant is also
invariant, since the determinant is invariant under the adjoint
rotation.   Thus the BGF generating functional $\tilde Z[J, {\cal
Q}]$ is invariant under (\ref{traOme}) and (\ref{traJ}).
\par
By using 
\begin{eqnarray}
 \tilde W[J, \Omega] := -i \ln \tilde Z[J, \Omega] ,
\end{eqnarray}
we define the background effective action
\begin{eqnarray}
 \tilde \Gamma[\tilde {\cal Q}, \Omega] := \tilde W[J, \Omega]
 - (J_\mu, \tilde Q_\mu) ,
\end{eqnarray}
where 
\begin{eqnarray}
 \tilde Q_\mu^A = {\delta \tilde W \over \delta J_\mu^A} .
\end{eqnarray}
From the invariance of $\tilde Z[J, {\cal Q}]$, it follows that 
$\tilde \Gamma[\tilde {\cal Q}, \Omega]$ 
is invariant under 
\begin{eqnarray}
 \delta \Omega_\mu 
 &=&  ({\cal D}_\mu[\Omega] \tilde \omega) ,
 \label{traOme2}
 \\
  \delta \tilde {\cal Q}_\mu  
  &=& i g[\tilde \omega, \tilde {\cal Q}_\mu]  ,
  \label{traQ2}
\end{eqnarray}
Since (\ref{traQ2}) is a homogeneous transformation, 
$\tilde \Gamma[0, \Omega]$ is invariant under the transformation
(\ref{traOme2}) alone.  Hence the effective action 
$\tilde \Gamma[0, \Omega]$ in the BGFM is an explicitly gauge
invariant functional of
$\Omega$, since (\ref{traOme2}) is just an ordinary gauge
transformation.  As a result, 1PI Green's functions generated by
differentiating 
$\tilde \Gamma[0, \Omega]$ with respect to $\Omega$ will obey the
naive Ward-Takahashi identities of gauge invariance.  Hence,
$\tilde \Gamma[0, \Omega]$ calculated in the BGFG is equal to the
conventional effective action
$\Gamma[\bar Q]$ 
with $\bar Q=\Omega$ calculated in an unconventional gauge which
depends on $\Omega$
\begin{eqnarray}
  \tilde F^A[{\cal Q}] := {\cal D}_\mu^{AB}[\Omega] 
  ({\cal Q}_\mu^B - \Omega_\mu^B) 
  = \partial_\mu {\cal Q}_\mu^A 
  + g f^{ABC} \Omega_\mu^B {\cal Q}_\mu^C 
  - \partial_\mu \Omega_\mu = 0  .
  \label{BGFgauge2}
\end{eqnarray}
Then we obtain
\begin{eqnarray}
 \tilde \Gamma[0, \Omega] = \Gamma[\bar Q]\big|_{\bar Q=\Omega} ,
\end{eqnarray}
as a special case of
\begin{eqnarray}
 \tilde \Gamma[\tilde Q, \Omega] = \Gamma[\bar Q]\big|_{\bar Q
= \tilde Q + \Omega} .
\end{eqnarray}
The 1PI Green functions calculated from the gauge invariant
effective action $\tilde \Gamma[0,\Omega]$ will be very different
from those calculated by conventional method in normal gauges. 
Nevertheless, the relation assures us that all gauge-invariant
physical quantities will come out the same in either approach
\cite{AGS83}.  Thus
$\tilde \Gamma[0, \Omega]$ can be used to generate the S-matrix of a
gauge theory in exactly the same way as the usual effective action
is employed.

\subsection{BRST version of the BGFM}
\par
Now we give the Becchi-Rouet-Stora-Tyupin (BRST) version of the
BGFM.  The BGF generating functional is rewritten into
\begin{eqnarray}
 \tilde Z[J, \Omega] 
 := \int [d{\cal Q}] [d\tilde C][d\bar {\tilde C}][d\tilde B]
 \exp \left\{ i S_{YM}[\Omega+{\cal Q}] 
 + i \tilde S_{GF}[{\cal Q}, \tilde C, \bar {\tilde C}, \tilde B]
 + i (J_\mu \cdot {\cal Q}_\mu)  
\right\} ,
\label{ZBGFM}
\end{eqnarray}
where $\tilde B$ is the auxiliary scalar field and $\tilde C, \bar
{\tilde C}$ are Hermitian anticommuting scalar field called the FP
ghost and  anti-ghost field,
$\tilde C^\dagger = \tilde C, 
\bar {\tilde C}^\dagger=\bar {\tilde C}$. Using the BRST
transformation, 
\begin{eqnarray}
   \tilde \delta_B \Omega_\mu(x)  
   &=& 0 ,
    \nonumber\\
   \tilde \delta_B {\cal Q}_\mu(x)  
   &=&  {\cal D}_\mu[\Omega+{\cal Q}] {\tilde C}(x)
   := \partial_\mu {\tilde C}(x) 
   - ig [\Omega_\mu(x)+{\cal Q}_\mu(x), {\tilde C}(x)],
    \nonumber\\
   \tilde \delta_B {\tilde C}(x)  
   &=& i{1 \over 2}g[{\tilde C}(x), {\tilde C}(x)],
    \nonumber\\
   \tilde \delta_B \bar {\tilde C}(x)  &=&   i \tilde B(x)  ,
    \nonumber\\
   \tilde \delta_B \tilde B(x)  &=&  0 ,
    \label{BRST0}
\end{eqnarray}
the gauge fixing and the FP ghost
terms for the BGG are combined into a compact form,
\begin{eqnarray}
  \tilde S_{GF}[{\cal Q}, \tilde C, \bar {\tilde C}, \tilde B]
  &:=& - \int d^Dx \ i  \tilde \delta_B \ {\rm tr}_G
  \left[ \bar {\tilde C}\left( \tilde F[{\cal Q}]+
  {\tilde \alpha \over 2}\tilde B \right) 
  \right] ,
  \label{GF0}
\end{eqnarray}
or
\begin{eqnarray}
  \tilde S_{GF}[{\cal Q}, \tilde C, \bar {\tilde C}, \tilde B]
  = \int d^Dx \  {\rm tr}_G \left[ 
  \tilde B  {\cal D}_\mu[\Omega] {\cal Q}_\mu 
  + {\tilde \alpha \over 2} \tilde B \tilde B
  + i \bar {\tilde C} {\cal D}_\mu[\Omega]
  {\cal D}_\mu[\Omega+{\cal Q}]\tilde C \right] ,
\end{eqnarray}
where $\tilde \alpha$ is the gauge-fixing parameter and $\tilde
\alpha=0$ corresponds to the Landau gauge (delta function gauge).
This is clearly BRST invariant
$\delta_B S_{GF}= 0$ due to nilpotency of the BRST transformation,
$\delta_B^2 \equiv 0$.
If the auxiliary field $\tilde B$ is integrated out, the gauge-fixing
part reads
\begin{eqnarray}
  \tilde S_{GF}[{\cal Q}, \tilde C, \bar {\tilde C}]
  = \int d^Dx \  {\rm tr}_G \left[  
    -{1 \over 2\tilde \alpha} ({\cal D}_\mu[\Omega] {\cal Q}_\mu)^2
  + i \bar {\tilde C} {\cal D}_\mu[\Omega]
  {\cal D}_\mu[\Omega+{\cal Q}]\tilde C \right] .
  \label{GFQ}
\end{eqnarray}
In fact, this recovers the original form (\ref{Z[J,O]}),  since
\begin{eqnarray}
  \det \left[ {\delta \tilde F^A \over \delta \tilde \omega^B}
\right]
= \int  [d\tilde C][d\bar {\tilde C}] \exp \left[  i\int d^Dx 
\  {\rm tr}_G \left( i \bar {\tilde C} {\cal D}_\mu[\Omega]
  {\cal D}_\mu[\Omega+{\cal Q}]\tilde C \right) \right] ,
\end{eqnarray}
The explicit form of the FP ghost term is
\begin{eqnarray}
 {\rm tr}_G \left[  i \bar {\tilde C} {\cal D}_\mu[\Omega]
   {\cal D}_\mu[\Omega+{\cal Q}]\tilde C  \right]
  &=& i \bar {\tilde C}^A [ \partial_\mu \partial_\mu \delta^{AB} 
  - g f^{ACB} {\uparrow \partial_\mu} (\Omega_\mu+{\cal Q}_\mu)^C
  + g f^{ACB} \Omega_\mu^B \partial_\mu 
  \nonumber\\&&
  + g^2 f^{ACE}f^{EDB} \Omega_\mu^C (\Omega_\mu+{\cal Q}_\mu)^D ]
  \tilde C^B ,
  \label{GFQ1}
\end{eqnarray}
and the gauge fixing term is
\begin{eqnarray}
 && {\rm tr}_G \left[  -{1 \over 2\tilde \alpha}
 ({\cal D}_\mu[\Omega] {\cal Q}_\mu)^2  \right]
 \nonumber\\
  &=&  -{1 \over 2\tilde \alpha} \left[ (\partial_\mu {\cal
Q}_\mu^A)^2 
  + 2 g f^{ABC} \Omega_\nu^B {\cal Q}_\nu^C \partial_\mu {\cal
Q}_\mu^A
  + g^2 f^{ABC} f^{ADE} \Omega_\mu^B {\cal Q}_\mu^C \Omega_\nu^D
{\cal Q}_\nu^E \right] .
  \label{GFQ2}
\end{eqnarray}
Feynmann rule for the BGFM is derived from the shifted action 
$S_{YM}[\Omega+{\cal Q}]$ and (\ref{GFQ}), see Abbott
\cite{Abbott82}.
In the limit $\Omega_\mu \rightarrow 0$, the BRST version of BGFM
reduces to the usual  BRST formulation of the Yang-Mills theory in
the Lorentz gauge, 
$F^A[Q] = \partial^\mu {\cal Q}_\mu$.
\par
The advantage of the BGFM becomes apparent when the
two-loop $\beta$ function is calculated.  The BGFM makes the
calculation much easier than previous calculations using the
conventional approach, see
\cite{Abbott81,AGS83}.

\section{Quantum theory of topological soliton and BGFM}
\setcounter{equation}{0}

\subsection{Summation over topological soliton background}
\par
In the conventional approach, the background field $\Omega_\mu$ is
chosen to be a solution of the classical field equation.  In
Yang-Mills theory, the equation of motion is given by
\begin{eqnarray}
  {\delta S_{YM}[{\cal A}] \over \delta {\cal A}_\mu^A} 
  \equiv {\cal D}^{AB}_{\nu} [{\cal A}]
  {\cal F}_{\mu\nu}^{B}[{\cal A}] = 0.
  \label{Feq}
\end{eqnarray}
Then, under the identification (\ref{separa}), 
\begin{eqnarray}
 {\cal A}_\mu = \Omega_\mu + {\cal Q}_\mu ,
\end{eqnarray}
the quantization is
performed around arbitrary but fixed background $\Omega_\mu$ which
satisfies (\ref{Feq}). In this paper, we consider the topologically
nontrivial field configuration as a background field
$\Omega_\mu$, around which the quantization of the Yang-Mills theory
is performed.   
Once a specific type of field configurations is chosen as the
background $\Omega_\mu$, we will include all possible 
configurations of the same type, in other words, we sum  up all
contributions coming from such a type of configurations.  
\footnote{
Such a procedure was performed so far in various forms, e.g., by
summing up the monopole-currents trajectories
\cite{ST78,BS78,AE99}.
}
Therefore,
in our formulation, a candidate for the generating functional of the
{\it total} Yang-Mills theory is given by
\begin{eqnarray}
 Z[J] &=& \int [d\Omega_\mu] \tilde Z[J, \Omega] 
 =: \int [d\Omega_\mu] \exp (i \tilde S_{eff}[J, \Omega]) ,
 \label{Z[J]0}
\end{eqnarray}
where we
have defined 
\begin{eqnarray}
 \tilde S_{eff}[J, \Omega] := - i \ln \tilde Z[J, \Omega] ,
\end{eqnarray}
and  $[d\Omega_\mu]$ is the integration measure specified later.
\par
Note that the action
$\tilde S_{eff}[J, \Omega]$ can have the local gauge invariance
(\ref{traOme}) for
$\Omega_\mu$ by virtue of the BGFM. 
Hence, the total Yang-Mills theory defined in this way is identified
with the (quantized) gauge theory with the action $\tilde
S_{eff}[J,\Omega]$,  provided that the integration measure $[d\Omega_\mu]$ is
gauge invariant.   However, in order to quantize the total Yang-Mills
theory correctly, we need to fix the local gauge invariance for the
non-Abelian gauge field
$\Omega_\mu$.  Thus, instead of (\ref{Z[J]0}), we define the
generating functional of the total Yang-Mills theory by
\begin{eqnarray}
 Z[J] = \int [d\Omega_\mu] 
 \delta(F^A[\Omega]) 
 \det \left[ {\delta F^A \over \delta \omega^B} \right]
 \tilde Z[J, \Omega] ,
\end{eqnarray}
or
\begin{eqnarray}
 Z[J] = \int [d\Omega_\mu] 
 \det \left[ {\delta F^A \over \delta \omega^B} \right]
 \exp \left(i \tilde S_{eff}[J, \Omega]
 -i{1 \over 2\alpha} 
 (F[\Omega] \cdot F[\Omega]) \right) ,
\end{eqnarray}
where the gauge fixing function $F^A$ is not necessarily equal
to the BGF gauge $\tilde F^A$.  The choice of $F^A$ is quite
important in our formulation for realizing topological soliton
background, as explained below.  
In order to be able to incorporate the topological soliton, the gauge
fixing function $F[\Omega]$ should be {\it nonlinear} in $\Omega$.
The measure $[d\Omega_\mu]$ must be chosen appropriately for
the topological soliton in question.  In the final stage the measure
is replaced by the integration over the
collective coordinates of the soliton.

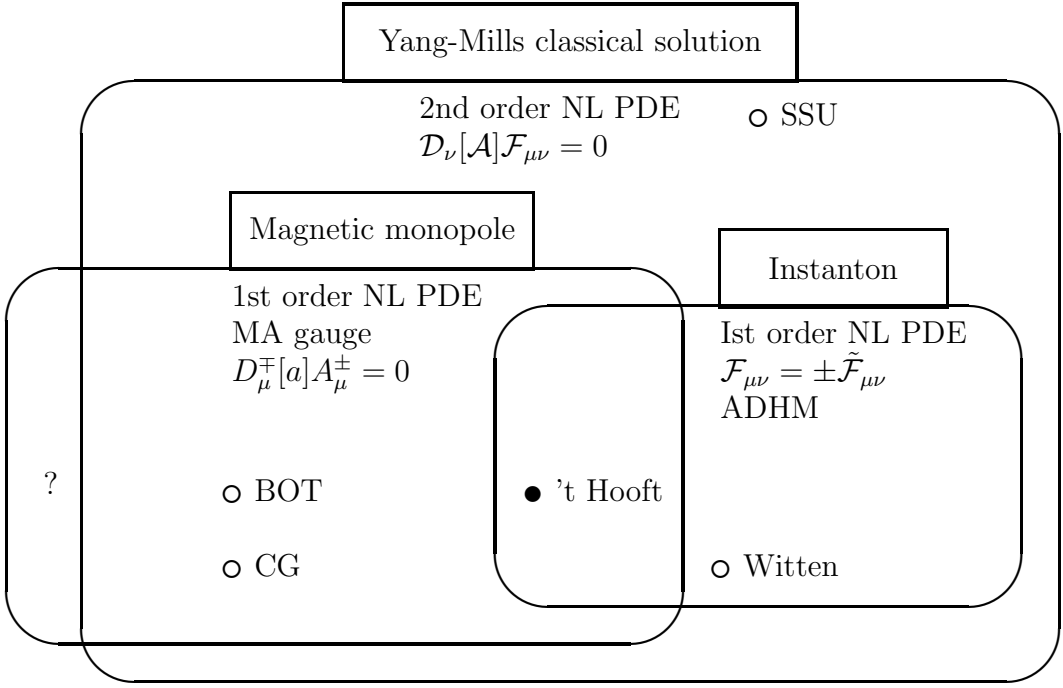
\begin{figure}
\begin{center}
\unitlength=1cm
 \begin{picture}(12,8)
\thicklines
 \put(3,7.5){\framebox(6,1){Yang-Mills classical solution}}
 \put(4,7.0){2nd order NL PDE}
 \put(4,6.5){${\cal D}_\nu[{\cal A}] {\cal F}_{\mu\nu}=0$}
 \put(6,3.5){\oval(13,8)}
 \put(8.5,7){\circle{0.2}}
 \put(8.8,6.9){SSU}
 \put(1.5,5){\framebox(4,1){Magnetic monopole}}
 \put(1.5,4.5){1st order NL PDE}
 \put(1.5,4.0){MA gauge}
 \put(1.5,3.5){$D^{\mp}_\mu[a]A_\mu^{\pm}=0$}
 \put(-1.0,2){\bf $?$}
 \put(3.0,2.5){\oval(9,5)}
 \put(5.5,2){\circle*{0.2}}
 \put(5.8,1.9){'t Hooft}
 \put(1.5,2){\circle{0.2}}
 \put(1.8,0.9){CG}
 \put(1.5,1){\circle{0.2}}
 \put(1.8,1.9){BOT}
 \put(8,4.5){\framebox(3,1){Instanton}}
 \put(8,4.0){Ist order NL PDE}
 \put(8,3.5){${\cal F}_{\mu\nu}=\pm \tilde {\cal F}_{\mu\nu}$}
 \put(8,3.0){ADHM}
 \put(8.5,2.5){\oval(7,4)}
 \put(8.0,1){\circle{0.2}}
 \put(8.3,0.9){Witten}
 \end{picture}
\end{center}
 \caption{
 Moduli space, i.e, space of solutions for the Yang-Mills
equation of motion (\ref{Feq}), self-dual instanton equation
(\ref{SDeq}) and the magnetic monopole equation (\ref{MAg0}) in the
MA gauge in four dimensions.  The instanton solution of the
self-dual equation (\ref{SDeq}) is also a solution of the Yang-Mills
equation of motion (\ref{Feq}).  The converse is not necessarily
true. In fact, the Sibner-Sibner-Uhlenbeck (SSU) solution 
\cite{SSU89} on
$S^4$ is a solution of the Yang-Mills field equation which is not a
solution of the self-dual equation
\cite{SSU89}.  The general instanton solution  on
$S^4$ can be constructed according to the Atiyah, Drinfeld, Hitchin
and Mannin (ADHM)
\cite{ADHM}.    
The explicit form for the multi-instanton is known in the specific
cases, e.g., 't Hooft type
\cite{Rajaraman89} or Witten type \cite{Witten77}.  Both types
include one-instanton solution of Belavin, Polyakov, Schwartz and
Tyupkin (BPST) \cite{BPST75}.  The multi-instanton solution of 't
Hooft type is also the solution of the magnetic monopole equation
(\ref{MAg}).   Some solutions are known for (\ref{MAg}),  Chernodub
and Gubarev (CG) \cite{CG95} and Brower, Orginos and Tan (BOT) 
\cite{BOT96}. See Appendix A.
The general solution of (\ref{MAg}) is not yet known. 
In principle, there may exist a solution of the monopole equation
which is not a solution of Yang-Mills field equation, indicated by
$?$ in the figure.  }
 \label{fig:moduli}
\end{figure}

\subsection{Yang-Mills instanton}
\par
In four-dimensional Euclidean space, the most popular
topologically nontrivial field configuration of pure Yang-Mills
theory is the instanton (anti-instanton) 
\cite{BPST75,tHooft76,Witten77,JNR77,ADHM,Coleman85,Rajaraman89}
which is a solution of the self-dual (self-antidual)
equation,
\begin{eqnarray}
  {\cal F}_{\mu\nu}[{\cal A}] = 
  \pm {\cal F}_{\mu\nu}^*[{\cal A}], \quad
  {\cal F}_{\mu\nu}^*[{\cal A}] := {1 \over 2}
  \epsilon_{\mu\nu\rho\sigma} {\cal F}_{\rho\sigma}[{\cal A}] ,
  \label{SDeq}
\end{eqnarray}
 with a finite action
\begin{eqnarray}
  S_{YM}[{\cal A}]  < \infty  .
\end{eqnarray}
The self-dual equation (\ref{SDeq}) is a first order nonlinear
partial differential equation (NL PDE), whereas the field equation
(\ref{Feq})is a second order nonlinear partial differential
equation.  
 The instanton is a kind of topological soliton which is possible
due to the nonlinearity of the self-dual equation.   Due to the
Bianchi identity,
\begin{eqnarray}
   {\cal D}_{\nu} [{\cal A}]{\cal F}_{\mu\nu}^*[{\cal A}] \equiv 0 , 
\end{eqnarray}
any instanton (anti-instanton) solution is also a solution of the
Yang-Mills field equation,
\begin{eqnarray}
  {\cal D}_{\nu} [{\cal A}]{\cal F}_{\mu\nu}[{\cal A}] = 0 ,
\end{eqnarray}
but the converse does not hold.
In fact,  the instanton
and the anti-instanton do not exhaust the solution of the Yang-Mills
field equation, since there exists at least one solution of the
Yang-Mills field equation (Sibner-Sibner-Uhlenbeck (SSU) solution
\cite{SSU89}) which is not a solution of the self-dual equation, see
Fig.~\ref{fig:moduli}. The existence of the instanton solution is
suggested from the non-triviality of Homotopy group
\cite{NS83,Nakahara90,Mermin79}
$\pi_3(G)$,
\begin{eqnarray}
 \pi_3(SU(N)) = {\bf Z} \ (N=2,3, \cdots) .
\end{eqnarray}
It is possible to construct the instanton background by choosing the
gauge fixing condition,
\begin{eqnarray}
 F^A[\Omega]  = {\cal F}_{\mu\nu}^{\pm}{}^A[\Omega],
 \quad 
 {\cal F}_{\mu\nu}^{\pm}{}^A[\Omega]
 := {\cal F}_{\mu\nu}^A[\Omega] \mp {\cal F}_{\mu\nu}^A{}^*[\Omega] .
 \label{SDYM}
\end{eqnarray}
It is shown \cite{KN98} that this choice leads to the topological
Yang-Mills theory \cite{Witten88,TQFT} which is an example of
the TQFT of Witten type.   Since the topological Yang-Mills theory is
derived from the N=2 Supersymmetric Yang-Mills theory by the
procedure called the twisting, this might shed more light on the
quark confinement based on the dual Meissner effect or the magnetic
monopole \cite{SW94}. However, quark confinement will be realized
only when the $N=2$ supersymmetry is broken down to $N=1$  by adding
the mass perturbation. 
Since we do not have any convincing argument to justify such a
scenario, we do not consider this possibility anymore in this
paper.

\subsection{Magnetic monopole current}
\par
In our formulation, however, the background field $\Omega_\mu$ is not
a priori required to be the classical solution of the field equation
(\ref{Feq}), when we consider the quantum theory of the background
field
$\Omega_\mu$. In quantum theory, it is not necessarily true that
the most dominant contribution is given
by the solution of the field equation.  This is obvious in the
functional integral approach because we must take into account the
entropy associated with the relevant field configurations, which
comes from the integration measure
$[d\Omega_\mu]$ of the functional integral.  In fact, whether the
phase transition occurs or not is determined according to the balance
between the action (energy) and the entropy, which is called the
action (energy)-entropy argument. What kind of field configuration is
important may vary from problem to problem.  
\par
In our approach, we take the magnetic monopole current as the
topologically nontrivial background $\Omega_\mu$.  
This choice is suggested from the recent result \cite{SY90} of
Monte Carlo simulations in lattice gauge theories; the (Abelian)
magnetic monopole after Abelian projection \cite{tHooft81} plays the
dominant role in quark confinement.  This fact is called the
(Abelian) magnetic monopole dominance
\cite{EI82}.
\par
The  magnetic monopole in pure Yang-Mills theory (without the
elementary Higgs scalar field) is obtained as follows.  First, we
restrict the Non-Abelian gauge group
$G$ to the subgroup $H$ ($G \rightarrow H$) and retain only the gauge
invariance for $H$, in other words, the gauge group element
$U(x) \in G$ is restricted to the coset $G/H$.  To obtain Abelian
magnetic monopole, $H$ is chosen to be the maximal
torus subgroup of $G$ (We will discuss other choices
in the final section).  
We realize this restriction by the partial gauge fixing.
The MAG is a  partial gauge
fixing so that $G/H$ is fixed and $H$ is retained 
by choosing
$F^A[\Omega]$ appropriately. The MA gauge condition is obtained by minimizing the ${\cal R}[{\cal
A}^U]$ with respect to the gauge rotation $U$ where
\begin{eqnarray}
{\cal R}[{\cal A}] := \int d^Dx \ {\rm tr}_{G/H}
 \left( {1 \over 2}{\cal A}_\mu(x) {\cal A}_\mu(x) \right) 
 \equiv  \int d^Dx \  
  {1 \over 2}A_\mu^a(x) A_\mu^a(x) ,
\end{eqnarray}
where we have used the Cartan decomposition which decomposes the
non-Abelian gauge field into the diagonal and the
off-diagonal pieces,
\begin{eqnarray}
  {\cal A}_\mu = {\cal A}_\mu^A T^A
  = a_\mu^\alpha T^\alpha + A_\mu^a T^a .
  \label{Cartandecomp}
\end{eqnarray}
Note that the trace is taken only on the coset part, see Appendix B.
A geometric meaning of this function is given in section 7.
According to the Cartan decomposition, the Abelian gauge potential is
defined by
\begin{eqnarray}
 a_\mu^\alpha(x) := {\rm tr}[{\cal H}_\alpha {\cal A}_\mu(x)] 
\end{eqnarray}
where  ${\cal H}_\alpha=T^\alpha (i=1, \cdots, {\rm rank}G)$ is the
Cartan subalgebra. 
For $G=SU(2)$, the differential MA gauge is obtained as
\begin{eqnarray}
  F^{a}[{\cal A}] := (\partial_\mu \delta^{ab} 
   - \epsilon^{ab3} A_\mu{}^3) A_\mu^b 
   := D_\mu{}^{ab}{}[A^3] A_\mu^b 
 \quad  (a,b = 1,2) ,
   \label{dMAG}
\end{eqnarray}
where $a_\mu=A_\mu^3$.
Note that the equation
\begin{eqnarray}
  D_\mu{}^{ab}{}[A^3] A_\mu^b = 0
 \quad  (a,b = 1,2) .
   \label{MAg0}
\end{eqnarray}
is a 1st order nonlinear partial differential equation. We call this
equation the monopole equation in what follows.
\par
Next, using the solution of $F^{a}[{\cal A}]=0$, 
the magnetic monopole current is defined by
\begin{eqnarray}
  k_\mu^\alpha := \partial_\nu \tilde f_{\mu\nu}^\alpha
  = {1 \over 2} \epsilon_{\mu\nu\rho\sigma} \partial_\nu
f_{\rho\sigma}^\alpha, 
\end{eqnarray}
where the Abelian field strength is given by
\begin{eqnarray}
  f_{\mu\nu}^\alpha  := \partial_\mu a_\nu^\alpha - \partial_\nu
a_\mu^\alpha . 
\end{eqnarray}
Due to the topological conservation law,
$
  \partial_\mu k_\mu^\alpha = 0 
$
the magnetic monopole current denotes a closed loop in four
dimensions.
The respective magnetic monopole is characterized by
an integer-valued topological (magnetic) charge 
$Q_m^\alpha := \int d^3x k_0^\alpha(x) \in {\bf Z}$
($\alpha=1,\cdots,N-1={\rm rank}SU(N)$).
For the static monopole, the monopole current is given by
$k_\mu^\alpha(x)=Q_m^\alpha \delta^3({\bf x}) \delta_{\mu 0}$
with $Q_m^\alpha$ being the magnetic charge.
\par
Finally, we must  check the finiteness of ${\cal R}[{\cal A}]$, which
is necessary to define the Morse function, see section 7.
The instanton solution give a finite Yang-Mills action, i.e., 
$S_{YM}[{\cal A}]<\infty$ irrespective of the gauge choice, as a
consequence of self-duality of the equation.
On the other hand, the magnetic
monopole solution of 
$F^{a}[{\cal A}]=0$ must give a finite 
${\cal R}[{\cal A}]$, i.e.,
${\cal R}[{\cal A}]<\infty$.  This condition leads to the finiteness
of the gauge-fixing action, 
\begin{eqnarray}
  S_{GF}[\Omega] < \infty ,
\end{eqnarray}
in the MA gauge.
In the gauge-fixed formulation of the quantum gauge field theory, the
gauge fixing part $S_{GF}$ is important as well as the Yang-Mills
action, $S_{YM}$.
In what folllows, it is very convenient to separate the
pure gauge piece in the gauge potential,
\begin{eqnarray}
  \tilde \Omega_\mu(x) 
  := {i \over g} \tilde U(x) \partial_\mu \tilde U^\dagger(x) 
  = \tilde \Omega_\mu^A(x) T^A , \quad \tilde U \in G/H .
\end{eqnarray}

\par
For $G=SU(2)$ and $H=U(1)$, it is shown that
the Abelian gauge potential calculated as
\begin{eqnarray}
  a_i(x) = {\rm tr}[{1 \over 2}\sigma_3 \tilde \Omega_i(x)]
(i=1,2,3)  
\end{eqnarray}
agrees exactly with the well-known static potential for the Dirac
magnetic monopole \cite{Dirac31,WY75,GO78}, see e.g. \cite{KondoI}.
\par
From the mathematical point of view, the existence of magnetic
monopole is consistent with the following relation for the Homotopy
group, 
\begin{eqnarray}
 \pi_2(G/H) = \pi_1(H) \ {\rm when} \  \pi_2(G)=0 .
\end{eqnarray}
Usually, the pure Yang-Mills theory does not have magnetic monopole
as a stable topological soliton.  This is consistent with 
\begin{eqnarray}
 \pi_2(G) = 0 .
\end{eqnarray}
Therefore, for the existence of the magnetic monopole in pure
Yang-Mills theory, the coset structure $G/H$ is an indispensable
ingredient.   For $G=SU(N)$, magnetic monopoles of
$N-1$ species are expected for the maximal torus group 
$H=U(1)^{N-1}$, since
\begin{eqnarray}
 \pi_2(SU(N)/U(1)^{N-1}) = \pi_1(U(1)^{N-1}) = {\bf Z}^{N-1} ,
\end{eqnarray}
whereas
\begin{eqnarray}
 \pi_2(SU(N)) = 0 \ (N=2,3, \cdots) .
\end{eqnarray}
\par

\section{Deformation of a topological field theory}
\setcounter{equation}{0}

\subsection{Change of field variables}
\par
For the decomposition of field variable,
\begin{eqnarray}
 {\cal A}_\mu(x)  = \Omega_\mu(x)   + {\cal Q}_\mu(x)  ,
 \label{deco}
\end{eqnarray}
it is possible to identify the gauge transformation 
\begin{eqnarray}
   \delta {\cal A}_\mu(x)  
   =  {\cal D}_\mu[A] {\omega}(x)
   := \partial_\mu {\omega}(x) - ig [{\cal A}_\mu(x),
{\omega}(x)],
\end{eqnarray}
with a set of transformations
\begin{eqnarray}
 \delta \Omega_\mu(x) 
 &=&  {\cal D}_\mu[\Omega] \omega(x) ,
 \label{traOme3}
 \\
  \delta {\cal Q}_\mu(x) &=& i g [\omega(x), {\cal
Q}_\mu(x)]  .
  \label{traQ3}
\end{eqnarray}
Here (\ref{traOme3}) and (\ref{traQ3}) correspond to (\ref{traOme}) and (\ref{traQ}) respectively.
Note that $\Omega_\mu$ transforms as an gauge field, while ${\cal
Q}_\mu$ as a adjoint matter field.
\par
When ${\cal A}_\mu$ is given by a finite gauge rotation (large gauge
transformation) $U(x)$ of
${\cal V}_\mu$, we take the following identification,
\begin{eqnarray}
  \Omega_\mu(x) := {i \over g} U(x) \partial_\mu U^\dagger(x), \quad
  {\cal Q}_\mu(x)  := U(x) {\cal V}_\mu(x) U^\dagger(x) ,
  \label{cov}
\end{eqnarray}
where we have identified $\Omega_\mu(x)$ with the background field
which is supposed to be generated from
$U(x)$.  This identification leads after simple calculation to 
\begin{eqnarray}
{\cal D}_\mu[\Omega] {\cal Q}_\mu
 &:=& \partial_\mu {\cal Q}_\mu 
 - i g[\Omega_\mu, {\cal Q}_\mu]  
 \\
 &=& \partial_\mu (U(x) {\cal V}_\mu(x) U^\dagger(x))
 + [  U(x) \partial_\mu U^\dagger(x), 
 U(x) {\cal V}_\mu(x) U^\dagger(x) ]  
 \nonumber\\
 &=&  U(x) \partial_\mu  {\cal V}_\mu(x) U^\dagger(x) .
 \label{gft}
\end{eqnarray}
Therefore, the BGF gauge for ${\cal Q}_\mu$,
\begin{eqnarray}
   {\cal D}_\mu[\Omega] {\cal Q}_\mu(x) = 0 ,
 \label{BGFG}
\end{eqnarray}
 is equivalent
to the Lorentz gauge for ${\cal V}_\mu$, 
\begin{eqnarray}
  \partial_\mu  {\cal V}_\mu(x) = 0 ,
  \label{Lorentz}
\end{eqnarray}
under the identification of the
variables (\ref{cov}).  
Under (\ref{cov}), we can rewrite (\ref{traOme3})  as
\begin{eqnarray}
 \delta U \partial_\mu U^\dagger 
 + U \partial_\mu \delta U^\dagger 
 = ig \omega U \partial_\mu U^\dagger 
 - ig U \partial_\mu(U^\dagger \omega) ,
\end{eqnarray}
and (\ref{traQ3}) as
\begin{eqnarray}
 \delta U {\cal V}_\mu U^\dagger 
 + U {\cal V}_\mu \delta U^\dagger 
 + U \delta {\cal V}_\mu U^\dagger
 = ig \omega U {\cal V}_\mu U^\dagger 
 - ig U {\cal V}_\mu U^\dagger \omega .
\end{eqnarray}
Therefore, in the BGF gauge, (\ref{traOme3}) and (\ref{traQ3}) reduce
to a set of transformations,
\begin{eqnarray}
 \delta U(x) 
 &=&  i g \omega(x) U(x) , \quad
 \delta U^\dagger(x)  =  - i g U^\dagger(x) \omega(x) ,
 \label{traU}
 \\
  \delta {\cal V}_\mu(x) &=& 0 ,
  \label{traV}
\end{eqnarray}
since the gauge degrees of freedom for ${\cal V}_\mu$ (small gauge
transformation) is fixed by the Lorentz gauge (\ref{Lorentz}). In
what follows, we assume that the non-compact gauge field variable
${\cal V}_\mu(x)$ does not have topologically nontrivial
configuration and all topologically nontrivial contributions come
from the compact gauge group  variable $U(x)$ alone.   We treat
${\cal V}_\mu(x)$ and $U(x)$ as if they are independent variables.
The topological soliton (magnetic monopole) is derived as a solution
of the nonlinear equation for
$\Omega_\mu$ which follows from the nonlinear gauge fixing condition
(MA gauge). The local gauge invariance of $\tilde Z[J,\Omega]$
written in terms of $\Omega_\mu$ and ${\cal Q}_\mu$ reduces to the
invariance   under the transformation (\ref{traU}), i.e. 
\begin{eqnarray}
  U(x) \rightarrow e^{ig\omega(x)}U(x) .
  \label{grot}
\end{eqnarray}
The measure $[d\Omega]$
invariant under (\ref{traOme}) is replaced by the invariant Haar
measure
$[dU]$ which is invariant under the local gauge rotation
(\ref{grot}).

\subsection{BRST formalism}

First, we rewrite the BRST formulation of BGFM in terms of new
variables.
By making the change of variable (\ref{cov}) which is a gauge
transformation of ${\cal V}(x)$ by $U(x)$, 
it turns out that the BRST transformation (\ref{BRST0}) for
the variables $\Omega_\mu, {\cal Q}_\mu, \bar {\tilde C}, \bar
{\tilde C}, \tilde B$ is rewritten into 
\begin{eqnarray}
   \tilde \delta_B U(x)  &=& 0 ,
    \nonumber\\
   \tilde \delta_B {\cal V}_\mu(x)  
   &=&  {\cal D}_\mu[{\cal V}] \gamma(x) ,
    \nonumber\\
   \tilde \delta_B \gamma(x)  
   &=& i{1 \over 2}g[\gamma(x), \gamma(x)],
    \nonumber\\
   \tilde \delta_B \bar \gamma(x)  &=&   i \beta(x)  ,
    \nonumber\\
   \tilde \delta_B \beta(x)  &=&  0 ,
    \label{BRST1}
\end{eqnarray}
where ${\cal V}_\mu, \gamma, \bar \gamma, \beta$ are the adjoint
rotation of 
${\cal Q}_\mu, \tilde C, \bar {\tilde C}, \bar B$ respectively,
\begin{eqnarray}
{\cal V}_\mu :=  U^\dagger {\cal Q}_\mu U, \quad
 \gamma := U^\dagger \tilde C U, \quad
 \bar \gamma := U^\dagger \bar {\tilde C} U, \quad
 \beta := U^\dagger \tilde B U  .
 \label{adjr}
\end{eqnarray}
Under the adjoint rotation (\ref{adjr}), the measure is invariant,
\begin{eqnarray}
[d{\cal V}] [d\gamma][d\bar \gamma][d\beta]
= [d{\cal Q}] [d\tilde C][d\bar {\tilde C}][d\tilde B] .
\end{eqnarray}
The Yang-Mills action is invariant 
\footnote{
For the definition of the field strength 
\begin{eqnarray}
  {\cal F}_{\mu\nu}[{\cal A}] 
  := \partial_\mu {\cal A}_\nu - \partial_\nu {\cal A}_\mu
   - i g [ {\cal A}_\mu,  {\cal A}_\nu] ,
\end{eqnarray}
the change of variable 
${\cal A}_\mu = U{\cal V}_\mu U^\dagger
+ { i \over g} U \partial_\mu U^\dagger$
leads to
\begin{eqnarray}
  {\cal F}_{\mu\nu}[{\cal A}] = 
  U {\cal F}_{\mu\nu}[{\cal V}] U^\dagger
  + { i \over g} U [\partial_\mu, \partial_\nu] U^\dagger .
\end{eqnarray}
Note that the second term 
${\cal F}_{\mu\nu}^s
:={ i \over g}U[\partial_\mu, \partial_\nu] U^\dagger$  
can have a nonzero value and may modify the action. If $U(x) \in G$
is restricted to the coset
$G/H$, it may yield a line-like singularity.  For example, it is
possible to have
${\cal F}_{xy}^s
:= {2\pi \over g} \delta(x)\delta(y)\theta(z) \sigma_3$
which corresponds to the presence of Dirac string extending into
the direction of the negative $z$ axis from the origin, see e.g.
Appendix C of \cite{KondoI}. Here the factor ${2\pi \over g}$
corresponds to the magnetic charge.  The same contribution as the
Dirac string can be incorporated by taking into account the magnetic
monopole instead of the Dirac string, as shown in
\cite{KondoI}.  
Moreover, the existence of such terms introduces rather singular
terms in the action.  Thus, we do not consider the effect of this
term in what follows. 
} 
under this change of variables,
\begin{eqnarray}
  S_{YM}[{\cal A}] = S_{YM}[\Omega+{\cal Q}]
  = S_{YM}[{\cal V}] .
\end{eqnarray}
The gauge fixing part (\ref{GF0}) for the BGF gauge is transformed
into  
\begin{eqnarray}
  \tilde S_{GF}[{\cal V}, \gamma, \bar \gamma, \beta]
  &:=& - \int d^Dx \ i  \tilde \delta_B \ {\rm tr}_G
  \left[ \bar \gamma \left( \partial_\mu {\cal V}_\mu +{\tilde \alpha
\over 2}\beta \right)   \right] 
\\
&=& \int d^Dx \ {\rm tr}_G \left[ \beta \partial_\mu {\cal V}_\mu
+ {\tilde \alpha \over 2} \beta \beta + i {\bar \gamma}
\partial_\mu {\cal D}_\mu[{\cal V}] \gamma 
\right] ,
\label{GF2}
\end{eqnarray}
where we have used (\ref{gft}) and (\ref{adjr}). It turns out that
the gauge fixing condition for
${\cal V}_\mu$ field is given by the Lorentz gauge
(\ref{Lorentz}).  Note that (\ref{GF2})  agrees with the form given
in \cite{KondoII}. This BRST transformation corresponds to the small
gauge transformation which does not change the topology of the gauge
field.
Thus the generating functional (\ref{ZBGFM}) is transformed as
\begin{eqnarray}
 \tilde Z[J, \Omega] 
 = \int [d{\cal V}] [d\gamma][d\bar \gamma][d\beta]
 \exp \left\{ i [ S_{YM}[{\cal V}] 
 + \tilde S_{GF}[{\cal V}, \gamma, \bar \gamma, \beta]
 + (J_\mu \cdot U {\cal V}_\mu U^\dagger) ]
\right\} 
\end{eqnarray}

\par
Next, we consider the total generating functional
\begin{eqnarray}
 && Z[J] 
 \nonumber\\&=& \int [d\Omega_\mu][dC][d\bar C][dB] 
 \tilde Z[J, \Omega] 
  \exp (i S_{GF}[\Omega, C, \bar C, B]) 
 \exp [i(J_\mu \cdot \Omega_\mu)]
 \\
 &=& \int [d\Omega_\mu] [dC][d\bar C][dB]
 \exp \{ i \tilde S_{eff}[J, \Omega]
 + i S_{GF}[\Omega, C, \bar C, B]
 +i(J_\mu \cdot \Omega_\mu) \} .
\end{eqnarray}
where we have introduced the source term $(J_\mu \cdot \Omega_\mu)$
for the background field. We introduce the BRST transformation,
\begin{eqnarray}
   \delta_B \Omega_\mu(x)  
   &=&  {\cal D}_\mu[\Omega] C(x)
   := \partial_\mu C(x) - i g[\Omega_\mu(x), C(x)],
    \nonumber\\
   \delta_B C(x)  
   &=& i{1 \over 2}g[C(x), C(x)],
    \nonumber\\
   \delta_B \bar C(x)  &=&   i B(x)  ,
    \nonumber\\
   \delta_B B(x)  &=&  0 ,
    \label{BRST2}
\end{eqnarray}
and the anti-BRST transformation, 
\begin{eqnarray}
   \bar \delta_B \Omega_\mu(x)  
   &=&  {\cal D}_\mu[\Omega] \bar C(x)
   := \partial_\mu \bar C(x) - i g[\Omega_\mu(x), \bar C(x)],
    \nonumber\\
   \bar \delta_B \bar C(x)  
   &=& i{1 \over 2}g[\bar C(x), \bar C(x)],
    \nonumber\\
   \bar \delta_B  C(x)  &=&   i \bar B(x)  ,
    \nonumber\\
   \bar \delta_B \bar B(x)  &=&  0 ,
    \label{BRST3}
\end{eqnarray}
where $\bar B$ is defined by
\begin{eqnarray}
 B(x) + \bar B(x) = g [C(x), \bar C(x)] .
\end{eqnarray}
The BRST and anti-BRST transformations have the following properties,
\begin{eqnarray}
 (\delta_B)^2 = 0, \quad (\bar \delta_B)^2 = 0, \quad
 \{ \delta_B,  \bar \delta_B \} 
 := \delta_B  \bar \delta_B + \bar \delta_B  \delta_B = 0 .
 \label{nilpotent}
\end{eqnarray}
\par
In what follows, we consider the variable $U(x)$ as the fundamental
variable instead of $\Omega_\mu(x)$. 
Then the BRST and anti-BRST transformations for $U$ and ${\cal
V}_\mu$ are given by
\begin{eqnarray}
    \delta_B U(x)  =  i g C(x) U(x),  \quad
   \bar \delta_B  U(x)  =   i g \bar C(x) U(x)  ,
\label{UBRST}
\end{eqnarray}
and
\begin{eqnarray}
 \delta_B {\cal V}_\mu(x) = 0 = \bar \delta_B {\cal V}_\mu(x) ,
 \label{VBRST}
\end{eqnarray}
which are the BRST version of (\ref{traU}) and (\ref{traV})
respectively. In fact, (\ref{UBRST}) reproduces the usual BRST
(\ref{BRST2}) and anti-BRST (\ref{BRST3}) transformations of the
gauge field,
\begin{eqnarray}
  \Omega_\mu(x) := {i \over g} U(x) \partial_\mu U^\dagger(x) .
\end{eqnarray}
Note that (\ref{UBRST}) and (\ref{VBRST}) lead to 
\begin{eqnarray}
 \delta_B \Omega_\mu(x) 
 &=&  {\cal D}_\mu[\Omega] C(x) ,
 \\
  \delta_B {\cal Q}_\mu(x) &=& i g [C(x), {\cal Q}_\mu(x)]   .
\end{eqnarray}
These are the BRST version of (\ref{traOme3}) and (\ref{traQ3})
respectively, since within the BGFM, 
$\Omega_\mu, C, \bar C, B$ or $U, C, \bar C, B$ are external
fields in the sense that they are not integrated out in the
measure $[d{\cal V}] [d\tilde C][d\bar {\tilde C}][d\tilde B]$.
Thus the generating functional of the total Yang-Mills theory reads
\begin{eqnarray}
 Z[J] =  \int [dU] [dC][d\bar C][dB]
 \exp \{ i \tilde S_{eff}[J, U]
 + i  S_{GF}[\Omega, C, \bar C, B]
 +i(J_\mu \cdot \Omega_\mu) \} ,
\end{eqnarray}
where $[dU]$ is the invariant Haar measure and we have redefined
\begin{eqnarray}
 \tilde S_{eff}[J, U] := - i \ln \tilde Z[J, \Omega] .
\end{eqnarray}

\par
In order to realize the magnetic monopole background, we adopt
the MA gauge for which the gauge fixing and the FP ghost terms are
written in the form \cite{KondoI}
\begin{eqnarray}
  S_{GF}[\Omega, C, \bar C, B]
  &:=& - \int d^Dx \ i  \delta_B \ {\rm tr}_{G/H}
  \left[ \bar C \left( F[\Omega]+{\alpha \over 2} B \right) 
  \right] ,
  \label{GF1}
\end{eqnarray}
where the trace is taken on the coset $G/H$, not the entire $G$.
For $G=SU(2)$, 
\begin{eqnarray}
  F^{a}[\Omega] := (\partial^\mu \delta^{ab} 
   - \epsilon^{ab3} \Omega^\mu{}^3) \Omega_\mu^b 
   := D^\mu{}^{ab}{}[\Omega^3] \Omega_\mu^b 
 \quad  (a,b = 1,2) .
   \label{dMAG2}
\end{eqnarray}
\par
 By adding an BRST-exact ghost self-interaction term, (\ref{GF1}) is cast into the more convenient form
\cite{KondoII},
\begin{eqnarray}
  S_{GF}'[\Omega, C, \bar C, B]
  := \int d^Dx \ i \delta_B \bar \delta_B 
  {\rm tr}_{G/H} \left[ {1 \over 2} \Omega_\mu(x) \Omega_\mu(x)
  - {\alpha \over 2} i C(x) \bar C(x) \right] .
  \label{GF'}
\end{eqnarray}
 From (\ref{nilpotent}),  $S_{GF}'$ is invariant under the  BRST and
anti-BRST transformations,
\begin{eqnarray}
 \delta_B S_{GF}' = 0 = \bar \delta_B S_{GF}' .
\end{eqnarray}
This action $S_{GF}'$ describes the topological soliton derived
from the nonlinear equation $F[\Omega]=0$, the monopole equation. 
This action is BRST exact and hence there is no local degrees of
freedom propagating in spacetime.  It describes the quantity related
to the global topology, as though the Chern-Simons theory describes
the linking of knots
\cite{Witten89}.  We call the theory with the
BRST exact action
$S_{GF}'=S_{TQFT}$ alone the topological quantum field theory (TQFT).
The generating functional is given by
\begin{eqnarray}
 Z_{TQFT}[J] =  \int [dU] [dC][d\bar C][dB]
 \exp \{  i S_{TQFT}[\Omega, C, \bar C, B]
 +i(J_\mu \cdot \Omega_\mu) \} .
 \label{TQFT}
\end{eqnarray}
In view of this, the above reformulation of the Yang-Mills theory was
called the deformation of the TQFT.
\footnote{ 
Different reformulations based on the similar idea have been
presented by many authors, e.g., by Hata and Taniguchi \cite{HT95},
and 
   Fucito, Martellini and Zeni \cite{FMZ97}.
}
\par
In the above rederivation, the fact that the field $\Omega_\mu$
behaves as if it is a gauge field ${\cal A}_\mu$ is essential.  This
is guaranteed by the BGFM.

\subsection{Expectation value}
\par
In our formulation,  an arbitrary function $f({\cal A})$ of
${\cal A}$  is written as
$f({\cal A}) = g({\cal V}_\mu, U) h(U)$ by making the change of
variable (\ref{deco}) and (\ref{cov}).
Then the expectation value is evaluated as
\begin{eqnarray}
  \langle f({\cal A}) \rangle_{YM} 
  = \langle \langle g({\cal V}_\mu, U) h(U) 
\rangle_{pYM}^{{\cal V}} \rangle_{TQFT}^U  
  = \langle \langle g({\cal V}_\mu, U)
\rangle_{pYM}^{{\cal V}} h(U) \rangle_{TQFT}^U .
\label{expec}
\end{eqnarray}
Here, taking the expectation value $\langle \cdot \rangle_{TQFT}^U$
corresponds to  summing over topological soliton contributions
by making use of the TQFT described by the variable $U$, whereas
$\langle \cdot \rangle_{pYM}^{{\cal V}}$ denotes the expectation
value for the deformation piece which is described by the usual
Yang-Mills theory with the variable
${\cal V}_\mu$ (Here $p$ denotes the perturbative).  
 Of course, we can change the
ordering of taking the expectation value, 
\begin{eqnarray}
  \langle f({\cal A}) \rangle_{YM} 
  = \langle \langle g({\cal V}_\mu,U) h(U) 
\rangle_{TQFT}^{U} \rangle_{pYM}^{{\cal V}} .
\end{eqnarray}
Both expressions should give the same result, if they are calculated
exactly.  
\par
Under the assumption of perturbative deformation, the expectation 
$\langle g({\cal V}_\mu,U) \rangle_{pYM}$ is calculated by expanding
the integrand $g({\cal V}_\mu,U)$ into power series in 
${\cal V}_\mu$.  
This is a minimal assumption in the practical calculation.
After that,
$\langle g({\cal V}_\mu,U) \rangle_{pYM}$ is still a function of $U$,
say, $p(U)$. Finally, the expectation 
$\langle p(U)h(U) \rangle_{TQFT}$ must be evaluated in the
non-perturbative way, since this piece estimates the soliton
contribution.   Perturbative deformation is an assumption that the
deformation part is evaluated in the perturbation theory in the
coupling constant
$g$.  In other words, all the essential non-perturbative
contributions are provided with the topological soliton described by
the TQFT.   
Actually, this strategy was performed in the
evaluation of the Wilson loop \cite{KondoII,KondoIV}.

\subsection{Abelian-projected effective gauge theory}
The above result should be compared with the previous formulation
\cite{QR97,KondoI} which begins with the generating
functional,
\begin{eqnarray}
 Z[J] = \int [d{\cal A}_\mu] [dC][d\bar C][dB]
 \exp \{ i S_{YM}[{\cal A}]
 + i S_{GF}[{\cal A}, C, \bar C, B]
 +i(J_\mu \cdot {\cal A}_\mu) \} .
\end{eqnarray}
First, following the Cartan decomposition (\ref{Cartandecomp}), the
non-Abelian gauge field was decomposed into the diagonal and the
off-diagonal pieces,
\begin{eqnarray}
  {\cal A}_\mu = {\cal A}_\mu^A T^A
  = a_\mu^i T^i + A_\mu^a T^a .
\end{eqnarray}
Then the MA gauge was imposed as a gauge fixing condition. 
Finally, all the off-diagonal fields taking values in the Lie
algebra of the coset
$G/H$ were integrated out in the functional integral,
\begin{eqnarray}
 Z[J] = \int [da_\mu^i] [dC^i][d\bar C^i][dB^i]
 \exp \{ i S_{diag}[a^i,C^i,\bar C^i,B^i] +i(J_\mu \cdot a_\mu)\} ,
\end{eqnarray}
where
\begin{eqnarray}
 && Z[a^i,C^i,\bar C^i,B^i]
 := \exp \{ i S_{diag}[a^i,C^i,\bar C^i,B^i] \} 
 \\
 &&:=  \int [dA_\mu^a] [dC^a][d\bar C^a][dB^a] 
 \exp \{ i S_{YM}[{\cal A}]
 + i S_{GF}[{\cal A}, C, \bar C, B]
 +i(J_\mu \cdot A_\mu) \}
\end{eqnarray}
The theory with the action 
$S_{diag}[a^i,C^i,\bar C^i,B^i]$ was called the Abelian-projected
effective gauge theory (APEGT).  
It has been shown \cite{QR97,KondoI} that the APEGT has the same beta
function as the original Yang-Mills theory, exhibiting the
asymptotic freedom, although the APEGT is an Abelian gauge theory.
\par
It turns out that the previous strategy presented in
\cite{QR97,KondoI} is equivalent to the above formulation presented in
this paper and that the results obtained in the previous works are
the immediate consequence of the present formulation, if we identify
the diagonal and off-diagonal fields with the background field and
the quantum fluctuation respectively, i.e.,
\begin{eqnarray}
  \Omega_\mu = a_\mu^i T^i,  \quad {\cal Q}_\mu = A_\mu^a T^a .
\end{eqnarray}
The theory with an action 
$S_{diag}[a^i,C^i,\bar C^i,B^i]$ is written in terms of only the
diagonal fields.  As long as the BGF gauge is imposed on the
off-diagonal field $A_\mu$, this theory becomes the Abelian gauge
theory, since the BGFM guarantees that the background field
$a_\mu$ transforms as a gauge field (Of course, the diagonal field
is reduced to the Abelian gauge field in this case).  Indeed, the BGF
gauge
$D^{ab}[a]A^b=0$ is nothing but the MA gauge.
Hence the coincidence of the beta function is understood from the
BGFM. 

\section{Strategy of a derivation of quark confinement}
\setcounter{equation}{0}

We consider the D-dim. QCD (QCD$_D$) with a gauge group G for $D >2$.
The (full) non-Abelian Wilson loop is defined as the path-ordered
exponent along a loop $C$,
\begin{eqnarray}
 W^C [{\cal A}] :=  {\rm tr} \left[ {\cal P} 
 \exp \left( i g \oint_C {\cal A}_\mu^A(x) T^A dx^\mu
 \right) \right] /{\rm tr}(1) .
\end{eqnarray}
We define the (full) string tension $\sigma$
by
\begin{eqnarray}
  \sigma := - \lim_{A(C) \rightarrow \infty}
  {1 \over A(C)} \ln \langle W^C[{\cal A}] \rangle ,
\end{eqnarray}
where $A(C)$ is the minimal area spanned by the Wilson loop $C$. 
The  non-zero string tension $\sigma \not=0$ implies
that the Wilson loop expectation value behaves for large loop as
\begin{eqnarray}
     \langle W^C[{\cal A}] \rangle 
    \sim \exp (- \sigma A(C)) .
\end{eqnarray}
This is called the area (decay) law.
The static potential $V(R)$ for a pair of quark and anti-quark  
is evaluated from the rectangular Wilson loop $C$ with sides $T$
 and $R$ ($A(C)=TR$) according to
\begin{eqnarray}
 V(R) =   - \lim_{T \rightarrow \infty} {1 \over T} \ln  
 \langle W^C[{\cal A}] \rangle .
\label{st}
\end{eqnarray}
The area law of the Wilson loop or non-zero string tension $\sigma
\not=0$ implies the existence of the linear part $\sigma R$ in the
static potential $V(R)$, leading to quark confinement. 

\par
In a series of papers \cite{KondoI,KondoII,KondoIII,KondoIV,KondoV},
a derivation of the area law of the Wilson loop in $QCD_4$ has been
given in the following steps.
\begin{enumerate}
\item[] Step 1: Reformulating the Yang-Mills theory as a deformation
of a TQFT in MA gauge \cite{KondoII}

\item[] Step 2: Parisi-Sourlas Dimensional reduction \cite{KondoII}

\item[] Step 3: Abelian magnetic monopole
dominance \cite{KondoIV}

\item[] Step 4: Instanton calculus \cite{KondoII} or large $N$ expansion \cite{KT99} 

\end{enumerate}
The first two steps are shown schematically as follows.
\par
\vskip 0.5cm
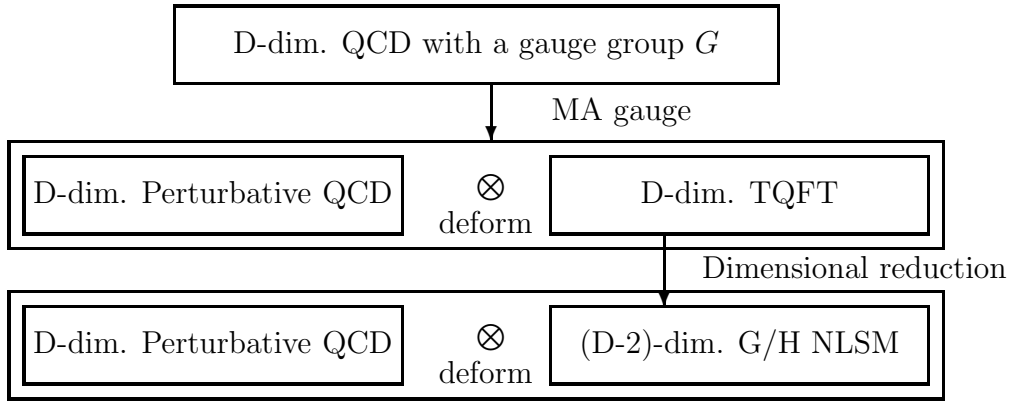
\begin{center}
\unitlength=1.0cm
\thicklines
 \begin{picture}(12,6)
 \put(2,5){\framebox(8,1){D-dim. QCD with a gauge group $G$}}
 \put(6.2,5){\vector(0,-1){0.8}}
 \put(7,4.5){MA gauge}
 \put(-0.2,2.8){\framebox(12.4,1.4){}}
 \put(0,3){\framebox(5,1){D-dim. Perturbative QCD}}
 \put(6,3.5){$\bigotimes$}
 \put(5.5,3){deform}
 \put(7,3){\framebox(5,1){D-dim. TQFT}}
 \put(8.5,3){\vector(0,-1){1}}
 \put(9,2.4){Dimensional reduction}
 \put(-0.2,0.8){\framebox(12.4,1.4){}}
 \put(0,1){\framebox(5,1){D-dim. Perturbative QCD}}
 \put(6,1.5){$\bigotimes$}
 \put(5.5,1){deform}
 \put(7,1){\framebox(5,1){(D-2)-dim. G/H NLSM}}
 \end{picture}
\end{center}

The following analyses within the above reformulation (the perturbative deformation of a TQFT) is based on an assumption that the contribution from ${\cal V}_\mu(x)$ can be treated in perturbation theory in the gauge coupling constant $g$.  This assumption leads to the topological sector dominance in the sense that the area law contribution comes solely from the contribution of $U(x)$ described by the TQFT and the remaining part ${\cal V}_\mu(x)$ does not contribute to the area law of the Wilson loop.

\subsection{Step 1: Reformulating the Yang-Mills theory as a
deformation of a TQFT in MA gauge}

QCD$_D$ is reformulated as a deformation of a TQFT$_D$ in
MA gauge.  The MA gauge is a partial gauge fixing such that the
coset part
$G/H$ of the gauge group $G$ is fixed and  the maximal torus group
$H$ is left as a residual gauge group.
\par
For  $G=SU(2)$, it has been shown \cite{KondoIV} that the expectation
value of the non-Abelian Wilson loop is rewritten using the
non-Abelian Stokes theorem
\cite{KondoIV} into 
\begin{eqnarray}
 && \langle W^C[{\cal A}] \rangle_{YM} 
\nonumber\\  
&=& \Biggr\langle \Biggr\langle 
  \exp \left[ i g J \oint_C dx^\mu n^A(x) {\cal V}_\mu^A(x) \right]
\Biggr\rangle_{pYM}
  \exp \left[ iJ \int_{S} d^2z \
\epsilon_{\mu\nu} {\bf n} \cdot (\partial_\mu {\bf n} \times
\partial_\nu {\bf n}) \right]
\Biggr\rangle_{TQFT} ,
\end{eqnarray}
where $S$ is a surface with a boundary $C$ ($\partial S=C$) and 
${\bf n}(x)=(n^1(x),n^2(x),n^3(x))$ is the three-dimensional unit
vector  (${\bf n}(x) \cdot {\bf n}(x) = 1$) defined by
\begin{eqnarray}
 n^A(x) T^A = U^\dagger(x) T^3 U(x) ,
 \quad T^A= {1 \over 2}\sigma^A \ (A=1,2,3) .
 \label{aop}
\end{eqnarray}
Here $J$ specifies the representation of the fermion in the
definition of the Wilson loop and 
$J=1/2$ corresponds to the fundamental representation.

\subsection{Step 2: Parisi-Sourlas dimensional reduction}

It has been shown \cite{KondoII} that TQFT$_D$ is equivalent to the
coset
$G/H$ nonlinear sigma model (NLSM) in (D-2) dimensions,
NLSM$_{D-2}$.   This is a consequence of Parisi-Sourlas dimensional
reduction \cite{PS79} due to the supersymmetry hidden in TQFT
(\ref{TQFT}). This is an advantage that we have chosen the MA
gauge.  
\par
As extensively discussed more than 20 years ago, 
QCD$_4$ and NLSM$_2$ have various common properties: 
renormalizability, asymptotic freedom (i.e., negative beta
function $\beta(g) < 0$), dynamical mass generation, existence of
instanton solution, no phase transition for any value of coupling
constant (i.e., one phase), etc. 
This similarity between two theories can be understood from this
correspondence,  
\begin{eqnarray}
 QCD_4 \supset TQFT_4 \Longleftrightarrow G/H~ NLSM_2 .
\end{eqnarray}
 For the $SU(N)$ gauge group, the existence of 2D instanton is guaranteed for any $N$, because
$\pi_2(SU(N)/U(1)^{N-1})=\pi_1(U(1)^{N-1})={\bf Z}^{N-1}$.  See the second paper in \cite{KT99} for details.
\par
For  $G=SU(2)$, G/H NLSM is nothing but the O(3) NLSM.
For the {\it planar} Wilson loop, the evaluation of the expectation
value 
$\langle \cdot \rangle_{TQFT}$ in  TQFT$_4$
\begin{eqnarray}
 \langle W^C[{\cal A}] \rangle_{YM} =
 Z_{TQFT_4}^{-1}\int [dU(x)]_{x \in {\bf R}^4} \exp (-S_{TQFT_4}[U] )
p(U)h(U)
\end{eqnarray}
is reduced to that in the coset $G/H$ NLSM$_2$
\begin{eqnarray}
\langle W^C[{\cal A}] \rangle_{YM} =
Z_{NLSM_2}^{-1}\int [d{\bf n}(x)]_{x \in {\bf R}^2} \exp
(-S_{NLSM_2}[{\bf n}] ) 
 p(U)h(U) ,
\label{Wc}
\end{eqnarray}
where we have used the notation (\ref{expec}) with
\begin{eqnarray}
 h(U) &:=& \exp \left[ iJ \int_{S} d^2z \
\epsilon_{\mu\nu} {\bf n} \cdot (\partial_\mu {\bf n} \times
\partial_\nu {\bf n}) \right] ,
\\
p(U) &:=& \Biggr\langle 
  \exp \left[ i g J \oint_C dx^\mu n^A(x) {\cal V}_\mu^A(x) \right]
\Biggr\rangle_{pYM} .
\end{eqnarray}

In the original Lagrangian of QCD, the scalar field is not included as an elementary field, but
it appears as a composite field according to (\ref{aop}). 
The unit vector ${\bf n}(x)$ plays the same role as the monopole
scalar field $\phi(x)$ which describes the 't Hooft-Polyakov
monopole \cite{tHooft74,Polyakov74}, for $G=SU(2)$, i.e.,
\begin{eqnarray}
 n^A(x) \leftrightarrow \hat \phi^A(x) := {\phi^A(x) \over |\phi(x)|}
 , \quad |\phi(x)| := \sqrt{\phi^A(x)\phi^A(x)} .
\end{eqnarray}

\subsection{Step 3: Abelian magnetic monopole dominance} 

The diagonal (or Abelian) string tension $\sigma_{Abel}$ is
defined by
\begin{eqnarray}
  \sigma_{Abel} 
   := -  \lim_{A(C) \rightarrow \infty} {1 \over A(C)} \ln  
  \left\langle  W^C[a^\Omega]
 \right\rangle_{TQFT_4} ,
\end{eqnarray}
by making use of the diagonal Wilson loop,
\begin{eqnarray}
  W^C[a^\Omega] &=&  \exp \left( i g J \oint_C 
  dx^\mu a_\mu^\Omega (x) \right),
\\
  \quad a_\mu^\Omega(x) &:=& \Omega_\mu^3(x) 
:= {\rm tr}(T^3 \Omega_\mu(x)) ,
\quad 
 \Omega_\mu(x) := {i \over g} U(x) \partial_\mu U(x)^\dagger .
  \label{dWl}
\end{eqnarray}
Owing to the dimensional reduction, we find
\begin{eqnarray}
 \left\langle W^C[a^\Omega] \right\rangle_{TQFT_4}
 =  \left\langle  W^C[a^\Omega]
 \right\rangle_{NLSM_2} ,
\end{eqnarray}
where
\begin{eqnarray}
\langle W^C[a^\Omega] \rangle_{NLSM_2} =
Z_{NLSM_2}^{-1}\int [d{\bf n}(x)]_{x \in {\bf R}^2} \exp
(-S_{NLSM_2}[{\bf n}] ) 
   W^C[a^\Omega] .
\label{Wc1}
\end{eqnarray}
Then it is shown that, in the limit of large Wilson loop, two string
tensions agree with each other,
$\sigma = \sigma_{Abel}$, since   
\begin{eqnarray}
 {1 \over A(C)} \left[  \ln  
  \left\langle W^C[{\cal A}] \right\rangle_{YM_4}
  - \ln \left\langle  W^C[a^\Omega] \right\rangle_{NLSM_2} \right] 
\downarrow 0 \quad (A(C) \uparrow \infty) ,
\end{eqnarray}
if we identify the deformation with the perturbative one. Thus, for the large planar (non-intersecting)  Wilson
loop , the full string tension $\sigma$ is saturated by the diagonal
string tension
$\sigma_{Abel}$.  This explains the Abelian dominance and magnetic
monopole dominance.
\par
This result is derived as follows.
In calculating (\ref{Wc}), if we put $p(U)\equiv 1$, then 
$\left\langle W^C[{\cal A}] \right\rangle_{YM_4}$ coincides with 
$\left\langle  W^C[a^\Omega] \right\rangle_{NLSM_2}$,  
since it is shown \cite{KondoII} that
$h(U) \equiv W^C[a^\Omega]$.  In our framework called the perturbative deformation of TQFT, $p(U)$ is estimated by making use of the power-series expansion in the coupling constant $g$ (or in the 't Hooft coupling $\lambda:= g^2N$ in the framework of large $N$ expansion, see \cite{KT99}) as  
\begin{eqnarray}
p(U) 
&=& 1 + \sum_{n=1}^{\infty} {(i g)^n \over n!} \Biggr\langle  \left(  J \oint_C dx^\mu n^A(x) {\cal V}_\mu^A(x) \right)^n \Biggr\rangle_{pYM} 
\nonumber\\
&=& 1 + \sum_{n=1}^{\infty} {(i gJ)^n \over n!}   \oint_C dx_1^{\mu_1} \cdots \oint_C dx_n^{\mu_n} n^{A_1}(x_1) \cdots n^{A_n}(x_n)  
\nonumber\\
&& \times
\langle  {\cal V}_{\mu_1}^{A_1}(x_1) \cdots {\cal V}_{\mu_n}^{A_n}(x_n) \rangle_{pYM} .
\end{eqnarray}
By calculating the expectation value
$\langle {\cal V}_{\mu_1}^{A_1}(x_1) \cdots {\cal V}_{\mu_n}^{A_n}(x_n) \rangle_{pYM}$, we can express $p(U)$ in terms of the ${\bf n}(x)$ fields which are defined on the loop $C$ embedded in the two-dimensional space.  However, it is shown that the additional contribution from $p(U)-1$ to the expectation value of the Wilson loop does not have the the area decay part.  Incidentally, although the perturbative deformation part is insufficient to derive the area law,  it leads to the running coupling constant which is governed by the renormalization group $\beta$ function of the original Ynag-Mills theory in consistent with the asymptotic freedom.
 
\subsection{Step 4: Instanton calculus}
\par
The whole problem is reduced to calculating the diagonal Wilson loop
in NLSM$_2$,
\begin{eqnarray}
 \left\langle  W^C[a^\Omega]
 \right\rangle_{NLSM_2} 
 = \left\langle    e^{i 2\pi J Q_S } 
 \right\rangle_{NLSM_2}  ,
\quad
  Q_S = {1 \over  8\pi } \int_S d^2z \ \epsilon_{\mu\nu}
{\bf n} \cdot (\partial_\mu {\bf n} \times \partial_\nu {\bf n})  .
\end{eqnarray}
Note that the integrand of $Q_S$ is the instanton density in
NLSM$_2$.  Therefore, $Q_S$ counts the number of instantons minus
that of anti-instantons inside the area $S(\subset {\bf R}^2)$
bounded by the Wilson loop $C$. 
This suggests that the quark confinement follows from the
condensation of topological soliton, the magnetic monopole.
\par
In this step we have employed the naive instanton calculus to
calculate the diagonal Wilson loop.  In the dilute gas approximation 
 the two-dimensional instanton contributions are summing
up according to 
\begin{eqnarray}
  \sum_{n_+=0}^{\infty} \sum_{n_-=0}^{\infty}
  {1 \over n_+! n_-!}
\int \prod_{i=1}^{n} d^2z_i \int \prod_{i=1}^{n} d\mu(\rho_i) 
\exp [-(n_++n_-)S_1(g) ]  e^{i 2\pi J Q_S } ,
\end{eqnarray}
where the action of NLSM$_2$ is replaced by
$(n_++n_-)S_1(g)$ using the numbers of
instanton and anti-instanton $n_+, n_-$ and the action for one
instanton $S_1(g)= 4\pi^2/g^2$ in NLSM$_2$. Thus the (infinite
dimensional) functional integral measure 
$[d{\bf n}(x)]_{x\in {\bf R}^2}$ has been replaced with the (finite
dimensional) integration with respect to the collective coordinates,
${z_i}$ (position of the instanton) and ${\rho_i}$ (size of the
instanton).  
Such reduction of degrees of freedom in the functional integration is
a common feature in TQFT as shown in  section 7.
\par
This leads to 
the area law of the diagonal Wilson loop and the
non-zero diagonal string tension
$\sigma_{Abel}$ for half odd integer $J$ 
or the fractional charge $q$.

\subsection{Area law and quark confinement}
\par
In the framework of the deformation of a TQFT for the Yang-Mills
theory, the non-zero string tension $\sigma$ in QCD$_4$ follows from
the non-zero diagonal string tension $\sigma_{Abel}$ in NLSM$_2$. 
The problem of proving area law in QCD$_4$ is reduced to the
corresponding problem in NLSM$_2$.

\par
All the above steps are exact except for the instanton
calculus of the Wilson loop in NLSM$_2$. 
For sufficiently large and planar Wilson loop, it was shown 
\cite{KondoII,KondoIV} that the string tension is given by
\begin{eqnarray}
 \sigma =   2B e^{-S_1} \left[ 1 - \cos 
\left( 2\pi J \right) \right]  ,
\quad S_1= {2\pi^2 \alpha \over g^2} ,
\label{str}
\end{eqnarray}
where $B$ is a constant with the mass-squared dimension coming from
the integration over the instanton size $\int d\mu(\rho)$ and
$S_1$ is the action for one instanton in NLSM$_2$.
\par
The result (\ref{str}) shows that for half odd integers
$J={1 \over 2}, {3 \over 2}, {5 \over 2}, \cdots$, the Wilson loop
exhibits area law for sufficiently large Wilson loop $C$, whereas 
the area law and the linear potential disappears for integers
$J=1,2,3, \cdots$.
Therefore, the fundamental fermion $J={1 \over 2}$ is confined, while
the adjoint fermion $J=1$ can not be confined.
\par
\par
In the above formulation using the MA gauge, it is the
{\it compact} residual Abelian group that plays the essential role in
evaluating the gauge invariant quantity.  This feature is very
similar to the situation in the lattice gauge theory.
In fact, the  result (\ref{str}) is a consequence of the periodicity
(or compactness) of the residual Abelian gauge group, i.e., maximal
torus group $U(1)$ of $SU(2)$, in the variable
$U$ after the MA gauge is chosen. 
On the other hand, the gauge degrees of freedom for the
non-compact field
${\cal V}_\mu$ have been completely fixed by the gauge fixing
condition of Lorentz type.  
The explicit expression (\ref{str}) depends on the approximation
taken in the instanton calculus, the periodicity of the string
tension (hence the absence of string tension for $J=1,2,\cdots$)
does not depend on the approximation.
\par
It is the perturbative part that gives the running of the coupling
constant $g$.  The running is governed by the renormalization group
beta function $\beta(g)$.  The usual Yang-Mills$_4$ theory
exhibits asymptotic freedom, e.g., for $G=SU(N_c)$ at one-loop
level,  
\begin{eqnarray}
     \beta(g) := \mu {dg(\mu) \over d\mu} 
     = - {b_0 \over 16\pi^2} g(\mu)^3 + \cdots ,
     \quad
     b_0 = {11N_c \over 3} > 0 .
\end{eqnarray}
In our framework, the correct beta function is derived based on the
BGFM, see \cite{QR97,KondoI}. 
For the static potential $V(R)$, the perturbative part gives a
Coulomb potential contribution 
$\alpha(\mu)/R$ where $\alpha(\mu):=g^2(\mu)/4\pi$ runs according to
the  $\beta(g)$.

\par
Similar strategy can also be applied to QED$_4$ ($G=U(1)$) to prove
the existence of strong coupling confinement phase \cite{KondoIII}. 
This follows from the existence of Berezinski-Kosterlitz-Thouless
transition of the O(2) NLSM$_2$.
The corresponding steps are shown as follows.

\par
\vskip 0.5cm
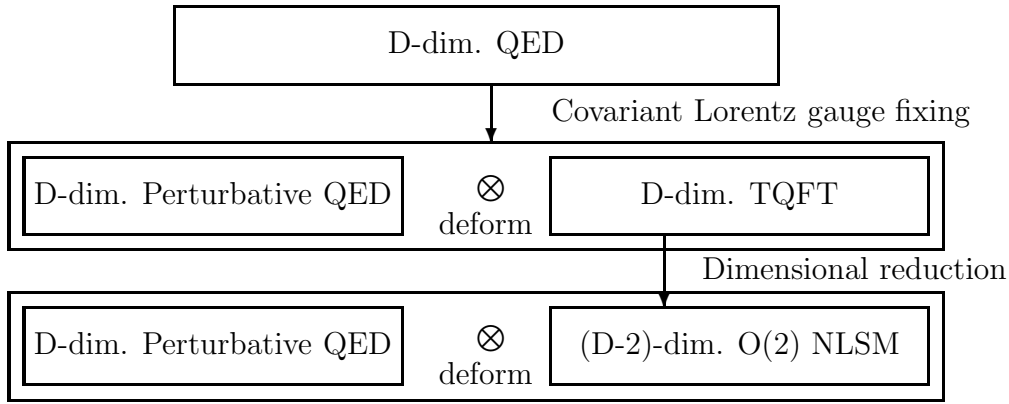
\begin{center}
\unitlength=1cm
\thicklines
 \begin{picture}(12,6)
 \put(2,5){\framebox(8,1){D-dim. QED}}
 \put(6.2,5){\vector(0,-1){0.8}}
 \put(7,4.5){Covariant Lorentz gauge fixing}
 \put(-0.2,2.8){\framebox(12.4,1.4){}}
 \put(0,3){\framebox(5,1){D-dim. Perturbative QED}}
 \put(6,3.5){$\bigotimes$}
 \put(5.5,3){deform}
 \put(7,3){\framebox(5,1){D-dim. TQFT}}
 \put(8.5,3){\vector(0,-1){1}}
 \put(9,2.4){Dimensional reduction}
 \put(-0.2,0.8){\framebox(12.4,1.4){}}
 \put(0,1){\framebox(5,1){D-dim. Perturbative QED}}
 \put(6,1.5){$\bigotimes$}
 \put(5.5,1){deform}
 \put(7,1){\framebox(5,1){(D-2)-dim. O(2) NLSM}}
 \end{picture}
\end{center}

This result enables us to give another derivation
of quark confinement in QCD based on the low-energy effective {\it
Abelian} gauge theory \cite{KondoI}, see \cite{KondoV}.
This viewpoint is more interesting in the sense that the
confinement-deconfinement transition can be discussed within the
same framework.

\subsection{Remarks and unresolved issues}
\par
The dilute gas approximation can be improved.
More systematic instanton
calculations enable us to identify an instanton solution with the
Coulomb gas of vortices
\cite{BL79,FFS79,BL81,Silvestrov90,Polyakov87}.
Consequently, the low-energy effective Abelian gauge theory belongs
to the strong coupling phase where the quark confinement is
realized, see \cite{KondoV}.
\par
The absence of intermediate Casimir scaling region (i.e.,
$\sigma=0$ for integer $J$) may be due to our simplified
treatment of  the instanton size.  In order to obtain
the result (\ref{str}) we have treated the instanton as if it is
exactly a point-like object in the dilute gas approximation.  The
Casimir scaling will be explained by taking into account the
size effect of the instanton, as performed for the center
vortex  by Greensite et al.
\cite{Greensite}.
\par
Recent investigations show that the QCD vacuum is a dual super
conductor caused by the condensation of magnetic monopole and that
the low-energy effective gauge theory is given the dual
Ginzburg-Landau theory.  However, numerical simulations claim that
the dual superconductor is near type I, rather than type II, see
\cite{Bali98}.  This result seems to contradicts with the analytical
studies.
\par
It is desirable to extend the above analyses into more general gauge
groups. The case of $G=SU(3)$ will be discussed in forthcoming
paper in detail.

\subsection{A proposal of numerical calculations}

Some of the implications from the above strategy will be
checked by direct numerical simulations on the lattice.
Due to difficulties of defining supersymmetry on the lattice, it
might be impossible to check directly the equivalence
between the 4D TQFT and the 2D NLSM. 
Nevertheless, it is desirable to check the
following statements:

\begin{enumerate}
\item
Validity of perturbative deformation of TQFT: The expectation
value of the diagonal Wilson loop in NLSM$_2$,
\begin{eqnarray}
 \left\langle  W^C[a^\Omega]
 \right\rangle_{NLSM_2} 
 = \left\langle    \exp \left[ i 2\pi J  
 {1 \over  8\pi } \int_S d^2z \ \epsilon_{\mu\nu}
{\bf n} \cdot (\partial_\mu {\bf n} \times \partial_\nu {\bf n})
 \right]
 \right\rangle_{NLSM_2}  ,
\end{eqnarray}
behaves as that of the non-Abelian Wilson loop in Yang-Mills$_4$,
\begin{eqnarray}
     \left\langle W^C[{\cal A}] \right\rangle_{YM_4} 
     =  \left\langle
     {\rm tr} \left[ {\cal P} 
 \exp \left( i g \oint_C {\cal A}_\mu^A(x) T^A dx^\mu
 \right) \right] \right\rangle_{YM_4} .
\end{eqnarray}
Two string tensions $\sigma_{Abel}$ and $\sigma$ agree with each
other.

\item
 Validity of instanton calculus: Only the instanton
contribution in NLSM$_2$ is sufficient to recover the Abelian string
tension,
$\sigma_{Abel}$.

\item
Existence of the scale:
The asymptotic scaling holds for the Abelian string tension
$\sigma_{Abel}$ calculated from the NLSM$_2$. 
\end{enumerate}
The results will prove or disprove validity of our strategy of
deriving quark confinement.   

\section{Gauge fixing and gluon mass}
\setcounter{equation}{0}

\subsection{A naive MA gauge}
A simple but ad hoc way to give the mass for the off-diagonal gluon
is to introduce the following mass term to the Yang-Mills action,
\begin{eqnarray}
   S_m = \int d^Dx {\rm tr}_{G/H} \left( {1 \over 2}m^2 
   {\cal A}_\mu {\cal A}_\mu \right) .
   \label{massterm}
\end{eqnarray}
This introduce the mass of the off-diagonal gluons in the tree level
and this explicitly break the gauge invariance corresponding to
$G/H$. Indeed, the mass term (\ref{massterm}) is derived as a gauge
fixing term as follows. The simplest MA gauge where the off-diagonal
part is made as small as possible will be the following gauge,
\begin{eqnarray}
  F^a[{\cal A}] := A_\mu^a = 0 .
  \label{Agauge}
\end{eqnarray}
In order to write the gauge-fixing action, we must introduce the
vector auxiliary field $B_\mu$ and the vector FP ghost $C_\mu$ and
anti-ghost $\bar C_\mu$, so that
\begin{eqnarray}
  S_{GF} = - \int d^Dx \ i\delta_B {\rm tr}_{G/H} \left[ \bar C_\mu
  \left( {\cal A}_\mu + {\alpha \over 2} B_\mu \right) \right] ,
\end{eqnarray}
where the nilpotent BRST transformation is constructed as
\footnote{
This BRST transformation does not leave the Yang-Mills action
invariant, unless the equation of motion is used.
So, it is unusual when we include the Yang-Mills action.
}
\begin{eqnarray}
   \delta_B {\cal A}_\mu &=& C_\mu,
   \nonumber\\
   \delta_B  C_\mu  &=& 0,
   \nonumber\\
   \delta_B \bar  C_\mu  &=& i B_\mu,
   \nonumber\\
   \delta_B B_\mu &=& 0 .
\end{eqnarray}
Eliminating the auxiliary field $B_\mu$, we reproduce the mass term,
\begin{eqnarray}
  S_{GF} =  \int d^4x  \ {\rm tr}_{G/H} \left[ 
   {m^2_\alpha \over 2} {\cal A}_\mu(x){\cal A}_\mu (x)
    + i  \bar  C_\mu(x) C_\mu(x) 
    \right] ,
\end{eqnarray}
where we have put $m^2_\alpha=1/\alpha$.
\footnote{
For $D=4$, the mass dimension is given as follows, dim[${\cal
A}_\mu$] = dim[$C_\mu$] = 1,  dim[$B_\mu$] = dim[$\bar  C_\mu$] = 3
and dim[$\alpha$]=-2.
}
In four-dimensions, the parameter $m_\alpha$ looks like a mass
which is arbitrary and can not be determined.
The BRST transformation is highly unusual, since it corresponds to
the gauge transformation much larger than the SU(N) gauge
transformation.  Note that $C_\mu^A$ has the same number of indices
as 
${\cal A}_\mu^A$.  Hence we can use $C_\mu^A$ to eliminate the fields
${\cal A}_\mu^A$ to obtain the vacuous theory.  The ghost field
$C_\mu^A$ has its own remaining ghost symmetry, parameterized by the
ghost field, $\phi^A$, the ghost for ghost, so the ghosts themselves
require more gauge fixing.
Note that the gauge fixing condition (\ref{Agauge}) does not allow
the topological soliton, since it is linear
in the field. 
\par
Making the change of variables with the adjoint orbit
parameterization
\begin{eqnarray}
 n^A(x)  = {\rm tr} \left[ U^\dagger(x) T^3 U(x)T^A \right] ,
 \quad T^A= {1 \over 2}\sigma^A \ (A=1,2,3)
 \label{aop2}
\end{eqnarray}
lead to the four-dimensional coset G/H
NLSM.
\begin{eqnarray}
  S_{GF} =  \int d^4x  \  \left[ 
 {m^2_\alpha \over 2} \partial_\mu {\bf n}(x)\partial_\mu {\bf n}(x)
    + \cdots
    \right] ,
\end{eqnarray}
since ${\cal A}_\mu= U\partial_\mu U^\dagger + \cdots$.
This is similar to a piece of the effective theory for the
low-energy QCD proposed by Faddeev and Niemi
\cite{FN98} based on Cho's works \cite{Cho80}.

\par

\subsection{The MA gauge}

The naive MA gauge above should be compared with the MA gauge.
The MA gauge 
\begin{eqnarray}
  F^{a}[\Omega] :=   (\partial^\mu \delta^{ab} 
   - \epsilon^{ab3} \Omega^\mu{}^3) \Omega_\mu^b 
   :=   D^\mu{}^{ab}{}[\Omega^3] \Omega_\mu^b 
 \quad  (a,b = 1,2) 
   \label{dMAG3}
\end{eqnarray}
is obtained by minimizing the ${\cal R}[{\cal A}^U]$
with respect to the gauge rotation $U$ where
\begin{eqnarray}
{\cal R}[{\cal A}] := \int d^Dx \ {\rm tr}_{G/H}
 \left( {k \over 2}{\cal A}_\mu(x) {\cal A}_\mu(x) \right) ,
\end{eqnarray}
where $k$ is a constant.
The MA gauge fixing leads to the gauge-fixing action (\ref{GF1})
where the gauge fixing parameter is arbitrary at this stage.
In our formulation, we demand the supersymmetry \cite{KondoII} of the
gauge fixing action.
Then the dimensional reduction \cite{PS79} occurs as a
spontaneous breaking of the supersymmetry (as explained below). 
This symmetry requirement has determined the form of the
gauge-fixing term (\ref{GF'}) and the result is independent from the
coefficient $k$ in
${\cal R}[{\ cal A}]$.
The explicit action (\ref{GF'}) after taking the BRST
transformation is rather complicated and does not have any apparent
mass term, see \cite{KondoI,KondoII}.  However, the dimensional
reduction
\cite{KondoII} leads to
\begin{eqnarray}
  S_{GF}' 
  &:=& \int d^{D-2}z \ 2\pi
  {\rm tr}_{G/H} \left[ {1 \over 2} {\cal A}_a(z,{\bf 0})
{\cal A}_a(z,{\bf 0})
  + i C(z,{\bf 0}) \bar C(z,{\bf 0}) \right]  
\\
&=& \int d^{D}x \ 
  {\rm tr}_{G/H} \left[ {2\pi \over 2} {\cal A}_a(x)
{\cal A}_a(x)
  + i 2\pi C(x) \bar C(x) \right]  \delta^2(\hat x) ,
  \label{GF''}
\end{eqnarray}
where $x=(z,\hat x) \in {\bf R}^D$ and $z \in {\bf R}^{D-2}$, $\hat x
\in {\bf R}^2$, $a=1, \cdots, D-2$. 
Hence, the MA gauge leads to the unusual mass term,
$m(x)=m(z,\hat x)=2\pi \delta^2(\hat x)$.  The mass is anisotropic
and the gauge field is massive only in $D-2$ dimensions. However, the
choice of the (D-2)-dimensional subspace is arbitrary. For $D=4$, the
equivalent action is given in the form of two-dimensional NLSM,
\begin{eqnarray}
  S_{GF} =  \int d^{2}z  \   \left[ 
 {2\pi/g^2 \over 2}
 \partial_a {\bf n}(z) \cdot \partial_a {\bf n}(z)
    + {\rm tr}_{G/H} \left( i 2 \pi \bar  C_\mu(z) C_\mu(z) \right)
    \right] .
\end{eqnarray}
It is known that the two-dimensional NLSM exhibits dynamical mass
generation, that is to say, the spectrum has a mass gap, although
the initial lagrangian does not have the usual mass term.  In this
sense, in the subspace
$R^2$ the gauge field can have the mass.
\footnote{
The mass generation due to dimensional reduction to the NLSM was
first demonstrated by Hata and Kugo \cite{HK85} in the context of color
confinement.  }

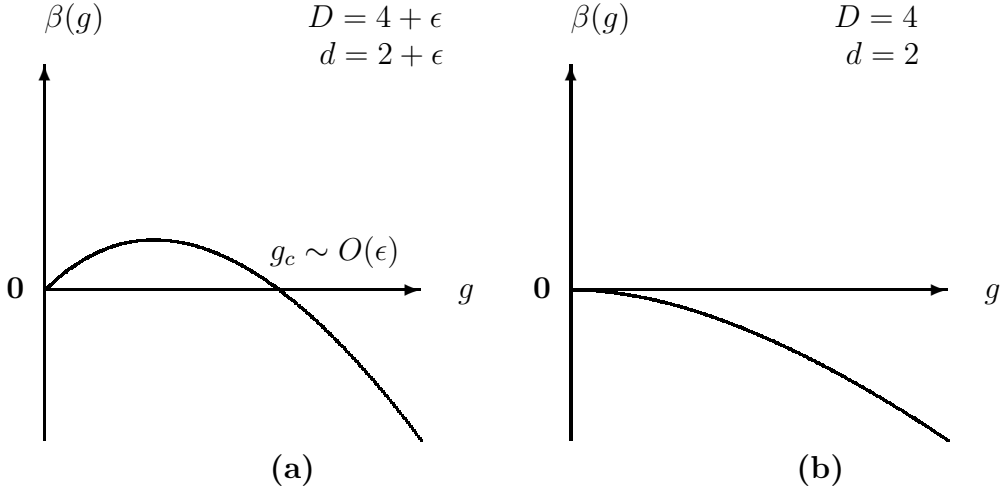
\begin{figure}
\begin{center}
\unitlength=1cm
 \begin{picture}(13,6)
  \put(3.5,5.5){\bf $D=4+\epsilon$}
  \put(3.5,5.0){\bf $~d=2+\epsilon$}
  \put(0,5.5){\bf $\beta(g)$}
  \put(3,2.4){\bf $g_c \sim O(\epsilon)$}
\thicklines
  \put(0,0){\vector(0,1){5}}
  \put(0,2){\vector(1,0){5}}
  \put(-0.5,1.9){\bf 0}
  \put(5.5,1.9){\bf $g$}
\bezier{300}(0,2)(2,4)(5,0)
  \put(3,-0.5){\bf (a)}
  \put(10.5,5.5){\bf $D=4$}
  \put(10.5,5.0){\bf $~d=2$}
  \put(7,5.5){\bf $\beta(g)$}
\thicklines
  \put(7,0){\vector(0,1){5}}
  \put(7,2){\vector(1,0){5}}
  \put(6.5,1.9){\bf 0}
  \put(12.5,1.9){\bf $g$}
\bezier{300}(7,2)(9,2)(12,0)
  \put(10,-0.5){\bf (b)}
 \end{picture}
\end{center}
 \caption{Renormalization group beta functions for the
$D$-dimensional Yang-Mills gauge theory and the $d$-dimensional NLSM
have the same form when $d=D-2$.
 (a) $D=4+\epsilon$ and $d=2+\epsilon$,
 (b) $D=4$ and $d=2$.}
 \label{fig:betafunc}
\end{figure}

\subsection{Spontaneous breakdown of hidden supersymmetry}
\par
A step of dimensional reduction is a little bit subtle.  In the
TQFT, the action is BRST exact by definition, so the partition
function and the expectation value of the gauge invariant operator do
not depend on the coupling constant.  However, the NLSM$_{D-2}$
obtained after the dimensional reduction from the TQFT$_D$ is not
topological and may depend on the coupling constant.  This seems at
first glance inconsistent.  
\par
This problem will be resolved as follows.  The dimensional reduction
is a consequence of the hidden supersymmetry in TQFT obtained in the
MA gauge, see \cite{KondoII}.  Here the supersymmetry implies the
invariance under the super rotation in the superspace 
$(x^\mu, \theta, \bar \theta)$, i.e., the orthsymplectic group
$OSp(D|2)$.  The advantage of introducing the superspace is to give a
geometric meaning to the BRST transformation.
In fact, the BRST symmetry becomes the translational invariance in
the superspace.  The BRST charges 
$Q_B$ and $\bar Q_B$ are generators of the translations in the direction of the
Grassmann variables
$\theta$ and $\bar \theta$, i.e., they are identified as
${\partial \over \partial \theta}$ and 
${\partial \over \partial \bar \theta}$, respectively.
In the process of the dimensional reduction we can choose
arbitrary two dimensions ${\bf R}^2$ from D dimensions $(x^1, \cdots,
x^D)\in {\bf R}^D$, since there is no privileged direction.  However,
once we have chosen specific two dimensions, the rotation symmetry
is partially broken by this procedure.   In this sense, the
dimensional reduction causes the  spontaneous breakdown of the
supersymmetry hidden in TQFT.  
\par
This becomes more clear in evaluating the expectation value of an
operator based on the dimensional reduction.  For this strategy to
work, the support of all the operators must be contained in the (D-2)
dimensional subspace to which the dimensional reduction occurs.  Such
an expectation value is obtained from the generating functional by
restricting the external source
${\cal J}(x,\theta,\bar \theta)$ to a (D-2)-dimensional
subspace  ${\cal J}((z,\hat x=0),\theta=0,\bar \theta=0)$ i.e., 
by putting $\hat x=\theta=\bar \theta=0$.  This is obtained as a consequence of restricting the orthosymplectic $OSp(D|2)$ rotation to the orthogonal $O(D-2)$ rotation, see section IV of the
paper
\cite{KondoII}.  
Of course, when a pair of quark and anti-quark exists,
it is convenient to choose the (D-2)-dimensional
subspace so that their trajectories are contained in the
subspace when $D\ge 4$.  
The supersymmetry (i.e., rotational invariance $OSp(D|2)$ in the BRST superspace) is broken into the orthogonal $O(D-2)$ rotation, whereas the BRST symmetries (i.e.,
translational invariances in the direction of $\theta, \bar \theta$ in the superspace)
is broken by putting $\hat x=\theta=\bar \theta=0$.
Note that the Hilbert space of the (D-2)-dimensional bosonic theory
is different from the original D-dimensional supersymmetric theory.
Consequently, (D-2)-dimensional bosonic theory is no
longer topological.
Thus, the NLSM$_2$ can be obtained without contradiction from TQFT$_4$ by dimensional reduction.
\par
Finally we consider the above result from a different point of
view.  We consider the
$4+\epsilon$ dimensional Yang-Mills theory and
$2+\epsilon$ dimensional NLSM.  For $\epsilon>0$, both theories have
two phases,  the disordered (high-temperature) phase in the strong
coupling region
$g>g_c$ and the ordered (low-temperature) phase in the weak coupling
region $g<g_c$ where $g_c \sim O(\epsilon)$.  The beta function is
expected to be positive for $0<g<g_c$ and negative for $g>g_c$.
See Fig.~\ref{fig:betafunc}(a).
As $\epsilon$ decreases, the ordered phases
shrinks and finally disappears   
($g_c(\epsilon) \downarrow 0$ as $\epsilon \downarrow 0$).  
In this limit, the beta
function becomes negative for any value of $g$, leading to the
asymptotic freedom for YM$_4$ and NLSM$_2$.
See Fig.~\ref{fig:betafunc}(b).
In this limit, the massless Nambu-Goldstone particle
associated to the spontaneous breaking of
$G$ to $H$ also disappear, as examined by Bardeen, Lee and Shrock
\cite{BLS76}.   
For $D=4$, thus, the massless Nambu-Goldstone (NG) particle associated
with the spontaneous breaking of supersymmetry, if any, can not
exist in two dimensions.
\footnote{
The above argument is clearly insufficient in the following sense.
The BRST symmetry is a global continuous symmetry.  If it is spontaneously broken,  the NG particle should appear, otherwise the Higgs mechanism should occur.
It has been shown that the massless NG boson in question does not appear in the physical state and that the dynamical Higgs mechanism does occur.  As a result, the off-diagonal gluons and the off-diagonal ghosts (anti-ghosts) become massive.  See the recent paper \cite{KS00} and references cited there.
}

\section{Geometric meaning of gauge fixing term}
\setcounter{equation}{0}

In quantizing the gauge theory, the procedure of gauge fixing is
indispensable to avoid infinities due to overcounting of gauge
equivalent configurations.  So in the quantized gauge theory we must
treat the gauge fixing term seriously as well as the gauge field
action.  Already at the level of classical theory, it is well known
that the gauge theory has a geometric meaning, i.e., gauge
theory is nothing but the geometry of connection.  In this section we
want to emphasize that the gauge fixing term may have a geometric
meaning from a viewpoint of global topology. 

\subsection{FP determinant}

The usual procedure of gauge fixing is to insert the identity
\begin{eqnarray}
  1 = \Delta_{FP}[{\cal A}] 
  \int [dU] \prod_{x} \delta (F^a[{\cal A}^U])  
\end{eqnarray}
into the functional integral 
\begin{eqnarray}
  Z = \int [d{\cal A}^U]  \exp (- S_{YM}[{\cal A}^U]) . 
\end{eqnarray}
Then we obtain
\begin{eqnarray}
  Z = \int [dU] \int [d{\cal A}^U] \Delta_{FP}[{\cal A}^U] 
   \prod_{x} \delta (F^a[{\cal A}^U]) 
  \exp (- S_{YM}[{\cal A}^U]) ,
  \label{Za}
\end{eqnarray}
since $\Delta_{FP}$ is gauge invariant, 
$\Delta_{FP}[{\cal A}] =\Delta_{FP}[{\cal A}^U]$.
The $\Delta_{FP}$ is calculated as follows.
\begin{eqnarray}
  \Delta_{FP}[{\cal A}]^{-1}
  &=&  \int [d\omega] \prod_{x} \delta (F^a[{\cal A}^\omega]) 
  \nonumber\\
  &=& \int [d\omega] \sum_{k} 
  {\delta \left( \omega - \omega_k \right) \over 
  | \det \left( {\delta F^a[{\cal A}^\omega] \over \delta \omega}
\right) |_{\omega=\omega_k} } 
\nonumber\\
&=& \sum_{k} {1 \over 
  | \det \left( {\delta F^a[{\cal A}^\omega] \over \delta \omega}
\right) |_{\omega=\omega_k} } .
\end{eqnarray}
When this result in the presence of Gribov copies is substituted into
(\ref{Za}), the BRST formulation does not work. Even when there is
no Gribov copies, we have the absolute value of the determinant,
\begin{eqnarray}
  \Delta_{FP}[{\cal A}]
  = \Big| \det \left( {\delta F^a[{\cal A}^\omega] \over \delta
\omega}
\right) \Big|_{\omega=\omega_k}  .
\end{eqnarray}
This expression is difficult to be used.
Therefore, we do not adopt this approach.  Rather we start from the
expression,
\begin{eqnarray}
  Z = \int [dU] \int [d{\cal A}^U]  
   \prod_{x} \delta (F^a[{\cal A}^U]) 
   \det \left( {\delta F^a[{\cal A}^\omega] \over \delta \omega}
\right)
  \exp (- S_{YM}[{\cal A}^U]) .
\end{eqnarray}
Such a formulation was proposed by Fujikawa \cite{Fujikawa79}.
We will show that such a proposal is very natural from the viewpoint
of global topology.

\subsection{Gauge fixing and global topology}

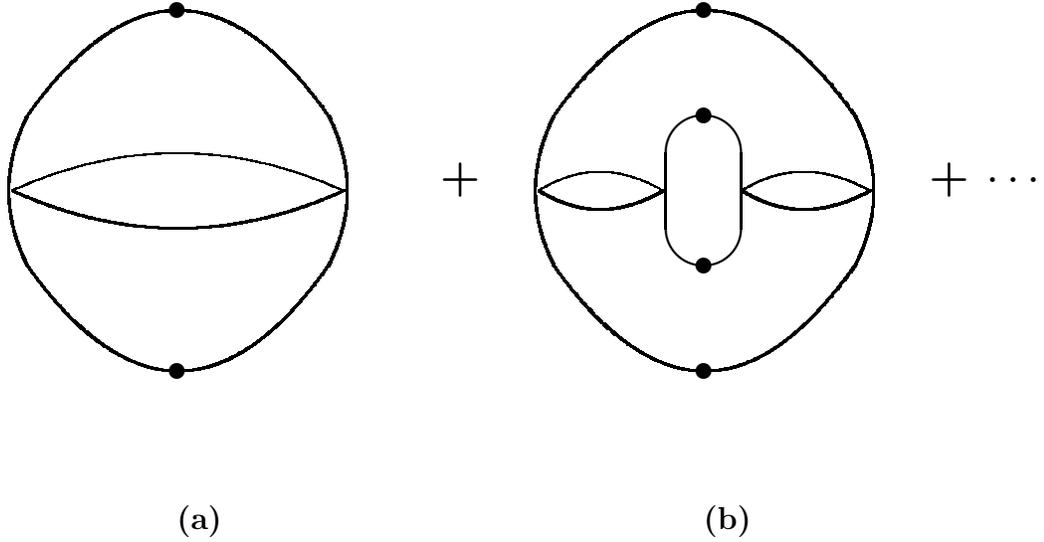
\begin{figure}
\begin{center}
\unitlength=1cm
 \begin{picture}(13,6)
\thicklines
\put(3,6.4){\circle*{0.2}}
\put(3,1.6){\circle*{0.2}}
\bezier{300}(1,5)(3,7.8)(5,5)
\bezier{300}(1,5)(0.5,4)(1,3)
\bezier{300}(5,5)(5.5,4)(5,3)
\bezier{300}(1,3)(3,0.2)(5,3)
\thinlines
\bezier{300}(0.8,4)(3,5)(5.2,4)
\thicklines
\bezier{300}(0.8,4)(3,3)(5.2,4)
\put(3,-0.5){\bf (a)}
\put(6.5,4){\Large \bf+}

\put(10,6.4){\circle*{0.2}}
\put(10,5.0){\circle*{0.2}}
\put(10,3.0){\circle*{0.2}}
\put(10,1.6){\circle*{0.2}}
\bezier{300}(8,5)(10,7.8)(12,5)
\bezier{300}(8,5)(7.5,4)(8,3)
\bezier{300}(12,5)(12.5,4)(12,3)
\bezier{300}(8,3)(10,0.2)(12,3)
\put(10,4){\oval(1,2)}
\thinlines
\bezier{100}(7.8,4)(8.6,4.5)(9.45,4)
\bezier{300}(10.5,4)(11.3,4.5)(12.2,4)
\thicklines
\bezier{300}(7.8,4)(8.6,3.5)(9.45,4)
\bezier{300}(10.5,4)(11.3,3.5)(12.2,4)
\put(10,-0.5){\bf (b)}

\put(13,4){\Large \bf+ $\cdots$}

 \end{picture}
\end{center}
 \caption{Various two-dimensional surfaces with different
topology. 
 The Euler number $\chi$ of the two-dimensional surface is determined
by the topology of the surface, i.e.,  $\chi=2-2g$ for the surface
with the genus $g$.  The Morse index $\nu_p=(-1)^{\lambda_p}$ is a
local quantity which is determined at the critical points (black dots in the
figures) of the
Morse function. The Euler number is equal to the sum of
the Morse indices  over all the critical points, i.e., 
$\chi=n_0-n_1+n_2$.  For example, it is easy to see that (a)
$\chi=2$ for the sphere $S^2$, (b) $\chi=0$ for the torus $T^2$.
}
 \label{fig:2Dsurface}
\end{figure}

\par
In the following, we use the notation 
$\phi^a = \{ {\cal A}_\mu^A(x) \}$ 
where $a$ denotes collectively $x,\mu,A$. 
If we take the gauge fixing condition
\begin{eqnarray}
  F^a[\phi] := {\delta f(\phi) \over \delta \phi^a} = 0,
\end{eqnarray}
the partition function in the gauge theory is given by
\begin{eqnarray}
  Z &:=&   \int [d\phi] \prod_{a} \delta (F^a[\phi])
  \det \left( {\delta F^a[\phi] \over \delta \phi^b}
\right)   e^{-S[\phi]}
\nonumber\\
 &=& \int [d\phi] \prod_{a} \delta 
  \left( {\delta f(\phi) \over \delta \phi^a} \right)
  \det \left( {\delta^2 f(\phi) \over \delta \phi^a \delta \phi^b}
\right)   e^{-S[\phi]} .
\end{eqnarray}
Let $M$ be the manifold $M$
of field configurations $\{ \phi^a \}$ and $f$ a continuous function
from $M$ to ${\bf R}$, $f:M \rightarrow {\bf R}$.
In order to see the geometric meaning of the gauge
fixing term, we consider the limit
$S[\phi]
\rightarrow 0$, i.e., only the gauge fixing and the corresponding FP
ghost term,
\begin{eqnarray}
  \chi_G := \int [d\phi] \prod_{a} \delta 
  \left( {\delta f(\phi) \over \delta \phi^a} \right)
  \det \left( {\delta^2 f(\phi) \over \delta \phi^a \delta \phi^b}
\right) .
\end{eqnarray}
Hence, from the property of the Dirac delta function, we have
\begin{eqnarray}
  \chi_G &=& \int [d\phi] \sum_{k} 
  {\delta \left( \phi - \phi_k \right) \over 
  | \det \left( {\delta^2 f(\phi) \over \delta \phi^a \delta \phi^b}
\right) |_{\phi=\phi_k} }
  \det \left( {\delta^2 f(\phi) \over \delta \phi^a \delta \phi^b}
\right)
  \\
  &=& \sum_{k: \nabla f(\phi_k)=0} {\rm sign} \left[
\det \left( {\delta^2 f(\phi) \over \delta \phi^a \delta \phi^b}
\right)_{\phi=\phi_k} \right] ,
\end{eqnarray}
where $\phi_k$ is a solution of $\nabla f(\phi)=0$ or
$F^a[\phi]=0$. Thus we obtain
\begin{eqnarray}
  \chi_G = \sum_{p: \nabla f(\phi_p)=0} \nu_p ,
\quad
\nu_p := {\rm sign} (\det H_f) = (- 1)^{\lambda_p} ,
  \label{Local}
\end{eqnarray}
where $H_f$ is the Hessian defined by
\begin{eqnarray}
  H_f := {\delta^2 f(\phi) \over \delta \phi^a \delta \phi^b} .
\end{eqnarray}
For a smooth function $f$, the Hessian is a symmetric matrix and
hence its eigenvalues are all real. The index $\lambda_P$ is equal to
the number of negative eigenvalues of the Hessian.   
Note that
$\chi$ is an integer.  In order to obtain this simple expression,
the existence of the FP determinant is indispensable.
The function $f$ is called the Morse function
\cite{Milnor63}, if all the critical points $P$ (i.e., $\nabla
f(P)=0$) of $f$ are non-degenerate, i.e.,
$\det(H_f)\not=0$.   For a finite dimensional case, it
is known that all non-degenerate critical points of $f$ are isolated
critical points, i.e., there is no other critical point in the
neighborhood of a critical point.   Whether the critical point is
degenerate or non-degenerate does not depend on how to choose the
coordinate systems in the manifold $M$ of field configurations $\{
\phi^a \}$. A convenient way to see the above situation is to use
the standard form. 
The quantity $\chi$ is obtained as the sum of the index $\nu_p$ over
all the critical points, once the function $f$ is given.  
See Fig.~\ref{fig:2Dsurface} for a simple case of
two-dimensional surface  $M_2$.  In that case, we have 
\begin{eqnarray}
  \chi(M_2) = n_0 - n_1 + n_2  ,
  \label{Euler1}
\end{eqnarray}
where $n_0, n_1, n_2$ are the total numbers of the minimal, saddle
and maximum points respectively.
\par
For a given field configuration 
$\phi^a$, we can consider the global topology.
The measure $[d\phi(x)]$ includes various field configurations with
 various global topology, each of which is characterized by an
appropriate topological invariant.  
By repeating the similar calculation, we obtain for the partition
function,
\begin{eqnarray}
  Z = \sum_{p: \nabla f(\phi_p)=0} \nu_p e^{-S[\phi_p]}
  = \sum_{p: \nabla f(\phi_p)=0} (- 1)^{\lambda_p} e^{-S[\phi_p]} .
  \label{Z}
\end{eqnarray}

\par
The two-dimensional case is rather simple, see
Fig.~\ref{fig:2Dsurface}. The Poincare'-Hopf theorem for the
two-dimensional surface states that $\chi$ defined in (\ref{Local})
is equal to the Euler number 
$\chi$ of the
two-dimensional surface $M_2$,
\begin{eqnarray}
  \chi(M_2)  = 2 - 2g ,
  \label{Euler2}
\end{eqnarray}
where  $g$ is the genus, i.e., number of handles. The Morse index
$\nu_p$ or
$\lambda_p$ is a local quantity, but, the sum $\chi$ given by
(\ref{Local}) is determined only from the global topology of the
surface
 (\ref{Euler2}) without any local information. 
 The Poincare-Hopf theorem gives a bridge
between the local geometry and global topology:
\begin{eqnarray}
  {\rm Local} \rightarrow \sum_{p} ({\rm local~index})_p = {\rm
topological~invariant} \leftarrow {\rm Global} .
\end{eqnarray}
This is a very important result for our purpose.  Because, by
deforming the surface in a continuous way, the location of the
critical point change and the Morse index at the new critical point
may also change, but the total sum of the Morse index is a
topological invariant which is determined only by the global
topology of the surface irrespective of the way of continuous
deformation.
\par
It is well known that the two-dimensional manifold is completely
classified by the genus or the Euler number.  This is not the case
in higher dimensions. In fact, we must treat the infinite
dimensional case. Even in the infinite dimensional case, for a
specific field configuration $M_\infty^\alpha$, a topological
invariant 
$Q(M_\infty^\alpha)$ will be determined.  Then it is expected
that the $\chi_G$ is expressed as a sum of topological invariants,
\begin{eqnarray}
  \chi_G = \sum_{Q} w_Q Q(M_\infty^\alpha)  .
\end{eqnarray}
In view of this, the functional we have chosen to derive the MA
gauge,
\begin{eqnarray}
{\cal R}[\Omega] := \int d^Dx \ {\rm tr}_{G/H}
 \left( {1 \over 2} \Omega_\mu(x) \Omega_\mu(x) \right)
\end{eqnarray}
is considered to be a Morse function.  
Indeed, the MA gauge condition is obtained as the gradient
of the Morse function ${\cal R}$ with respect to the large gauge
transformation,
\begin{eqnarray}
  {\delta {\cal R}[\Omega^\omega] \over \delta \omega} 
  = F[\Omega] .
\end{eqnarray}
The topology change in the field configurations may be caused by the
large (or finite) gauge transformation allowed in the
measure $[dU]$, since the measure is invariant under the
global gauge rotation 
$U \rightarrow e^{i\omega}U$. 
Thus, the gauge fixing and the associated FP term
when integrated out by the functional measure $[d\phi]$ can have a
geometric meaning which is related to the global topology of the
field configurations.  The Morse function is a tool of probing the
global topology allowed in the functional space of the
field configurations by gathering the local information at all the
critical points.  
The partition function of Yang-Mills theory is
given by  
\begin{eqnarray}
  Z_{YM}[0] = \sum_{p: F(\Omega_p)=0} \nu_p e^{-S_{eff}[\Omega_p]}
  = \sum_{p: F(\Omega_p)=0} (- 1)^{\lambda_p}
e^{-S_{eff}[\Omega_p]} .
  \label{Z0}
\end{eqnarray}

\par
Finally, we consider how $\chi$ changes when we change the Morse
function
$f$.
In two-dimensional case, $\chi(M_2)$ is determined by the topology of
manifold $M_2$ irrespective of the choice of Morse function $f$.
If this feature survives in the infinite dimensional case, we can
conclude that the global topology of the Yang-Mills theory does not
depend on the way of gauge fixing.

\subsection{Morse function and BRST transformation}

By introducing the auxiliary field $B$ and the FP ghost and
anti-ghost fields, $\psi, \bar \rho$, we can write
\begin{eqnarray}
  \chi :=   \int [d\phi] \prod_{a} \delta (F^a[\phi])
  \det \left( {\delta F^a[\phi] \over \delta \phi^b}
\right)   
= \int [d\phi] [dB] [d\psi] [d\bar \rho] e^{-S_{GF}[\phi]} ,
\end{eqnarray}
where the gauge-fixing action reads
\begin{eqnarray}
   S_{GF} = \int d^Dx \left[ {\alpha \over 2} B^2 + BF
   - \bar \rho F'[\phi] \psi \right]
   = \int d^Dx \ \delta_B \left[ \bar \rho 
   \left( F[\phi]+{\alpha \over 2} B \right) \right]
\end{eqnarray}
with the nilpotent BRST transformation,
\begin{eqnarray}
   \delta_B \phi &=& \psi,
   \nonumber\\
   \delta_B \psi &=& 0,
   \\
   \delta_B \bar \rho &=& B,
   \\
   \delta_B B &=& 0 .
\end{eqnarray}
By eliminating the auxiliary field $B$, we have
\begin{eqnarray}
  \chi &=& \int [d\phi] [d\psi] [d\bar \rho] e^{-S_{GF}'},
  \\
  S_{GF}' &=& \int d^Dx \left[ -{1 \over 2\alpha} (F[\phi])^2 
   - \bar \rho F'[\phi] \psi \right] ,
\end{eqnarray}
where 
\begin{eqnarray}
   \delta_B \phi &=& \psi,
   \nonumber\\
   \delta_B \psi &=& 0,
   \\
   \delta_B \bar \rho &=& {F[\phi] \over \alpha} .
\end{eqnarray}
The critical point  
$F[\phi] := {\partial f(\phi) \over \partial \phi}= 0$
corresponds to the fixed point of BRST transformation.
Therefore, the integration $\int [d\phi] \delta(F[\phi]) \cdots$
is localized on the fixed point of BRST transformation.  This is a
characteristic feature of topological quantum field theory.
The condition $F^a[\phi]=0$ is regarded as a non-linear partial
differential equation.  The space of parameters characterizing
the solution of this equation is called the moduli space.  In the
TQFT, the above argument shows that the infinite dimensional
functional integral reduces to finite dimensional integral on moduli
space.  For example, for the Yang-Mills instanton with $Q=k$, ${\rm
dim}M=8k < \infty$.

\section{Conclusion and discussion}
\setcounter{equation}{0}

In this paper we have derived a reformulation of the Yang-Mills
theory based on the background field method.  The reformulation
identifies the Yang-Mills theory as a deformation of a topological
quantum field theory as proposed in \cite{KondoII}.  The background
field is given by a topological soliton.  
\par
In order to show quark confinement, the condensation of a topological
soliton is necessary to occur.  This has been actually derived by
summing up the topological soliton contributions, provided that the
topological soliton is described by the topological field theory. The
topological field theory has been derived from the gauge fixing term
corresponding to the {\it nonlinear} gauge fixing condition, the
maximal Abelian gauge. 
The maximal Abelian gauge implies that the topological soliton in
question is nothing but the magnetic monopole current, the
four-dimensional version of the magnetic monopole. 
The result ensures that the quark confinement is realized in the QCD
vacuum as a dual superconductor.
Furthermore, we have proposed a numerical simulation which is able to
confirm the validity of the above reformulation.
\par
We have discussed a novel mechanism for the mass generation for the
gauge field, i.e., dynamical mass generation as the
dimensional reduction which causes the spontaneous breakdown of the
hidden supersymmetry in the
topological field theory. Moreover, we have suggested that the gauge
fixing action may have the geometric meaning from the view point
of global topology by making use of the Morse function.
\par
In this paper we have restricted our consideration to the maximal
Abelian gauge where the residual gauge group $H$ is the maximal
torus group of the non-Abelian gauge group $G$
($H=U(1)^{N-1}$ for $G=SU(N)$), although our formulation can be
applied to any choice of $H$.  Therefore the topological soliton is
given by the Abelian magnetic monopole.  However, it is possible to
consider other choices for the residual gauge group $H$
(especially for $G=SU(N) (N \ge 3)$) which leads to the topological
soliton other than the Abelian magnetic monopole, e.g., non-Abelian
magnetic monopole, center vortex. Either choice will lead to the
quark confinement. From the viewpoint of {\it color} confinement,
however, the maximal torus group for $H$ is not necessarily the
best choice.  The details will be given in the subsequent paper
\cite{Kondo99}.

\appendix
\section{Overlapping between monopole current and instanton
solutions}

The MA gauge (\ref{dMAG}) is written as
\begin{eqnarray}
  (\partial_\mu \mp i A_\mu^3) A_\mu^{\pm} = 0 ,
  \label{MAg}
\end{eqnarray}
where
\begin{eqnarray}
  A_\mu^{\pm} := {1 \over \sqrt{2}}(A_\mu^1 \pm i A_\mu^2) .
\end{eqnarray}
We show that the gauge potential of the form,
\begin{eqnarray}
  {\cal A}_\mu^A(x) = \eta_{\mu\nu}^A \partial_\nu f(x) ,
  \label{ansatz}
\end{eqnarray}
satisfies the monopole equation (\ref{MAg}) for {\it arbitrary}
function $f$ as long as 
$[\partial_\mu, \partial_\nu] f = 0$.
Here the $\eta$-symbol ('t Hooft symbol) is defined by
\begin{eqnarray}
   \eta_{\mu\nu}^A := \epsilon_{A\mu\nu}
  + \delta_{A\mu} \delta_{4\nu} - \delta_{A\nu}\delta_{4\mu} ,
\end{eqnarray}
for $\mu,\nu=1,2,3,4$, $A=1,2,3$ (we have assumed 
$\epsilon_{A4\nu}=\epsilon_{A\mu4}=0$). 
\par
Substituting the ansatz (\ref{ansatz}) into the definition of
$A_\mu^{\pm}$, we have
\begin{eqnarray}
   A_\mu^3 A_\mu^{\pm} = {1 \over \sqrt{2}}
  \eta^3_{\mu\rho} (\eta^1_{\mu\sigma} \pm i \eta^2_{\mu\sigma})
   \partial_\rho f \partial_\sigma f
= {1 \over \sqrt{2}} 
   (\eta^2_{\rho\sigma} \mp i\eta^1_{\rho\sigma}) 
\partial_\rho f \partial_\sigma f = 0,
\end{eqnarray}
where we have used  the relations \cite{tHooft76}
\begin{eqnarray}
   \eta^A_{\mu\rho} \eta^B_{\mu\sigma}
   = \delta_{AB} \delta_{\rho\sigma} 
   + \epsilon_{ABC} \eta^C_{\rho\sigma} ,
   \\
   \eta_{\mu\nu}^A=-\eta_{\nu\mu}^A .
\end{eqnarray}
On the other hand, we find
\begin{eqnarray}
  \partial_\mu {\cal A}_\mu^A(x) 
  = \eta_{\mu\nu}^A \partial_\mu \partial_\nu f(x) ,
\end{eqnarray}
and hence
\begin{eqnarray}
  \partial_\mu  A_\mu^{\pm} := {1 \over \sqrt{2}}
  (\partial_\mu  A_\mu^1 \pm i \partial_\mu A_\mu^2) 
  = {1 \over \sqrt{2}}(
  \eta_{\mu\nu}^1 \partial_\mu \partial_\nu f
\pm i \eta_{\mu\nu}^2 \partial_\mu \partial_\nu f)  .
\end{eqnarray}
Note that
$\eta_{\mu\nu}^A \partial_\mu \partial_\nu f(x)
= {1 \over 2}\eta_{\mu\nu}^A [\partial_\mu, \partial_\nu] f(x)$.
Therefore, the MA gauge (\ref{MAg}) is satisfied for any function
$f$ as long as 
$[\partial_\mu, \partial_\nu] f = 0$.
\par
The ansatz (\ref{ansatz}) is the same as the multi-instanton
solution of 't Hooft type.
The $\eta$-symbols are self-dual in the vector indices,
\begin{eqnarray}
   \eta_{\mu\nu}^A  = {1 \over 2} \epsilon_{\mu\nu\alpha\beta}
   \eta_{\alpha\beta}^A .
\end{eqnarray}
The instanton solution is obtained assuming 
$[\partial_\mu, \partial_\nu] f = 0$, since the instanton is the
point defect in four dimensions.  For the magnetic monopole current
(the line defect) in four dimensions,  $[\partial_\mu, \partial_\nu]
f \not= 0$  can happen, see e.g. Appendix C of \cite{KondoI}.  Such a
possibility has not been studied so far.
\par
The $n$-instanton solution in the singular gauge is given by   
\begin{eqnarray}
  f(x) = \ln \left[ 1+ \sum_{k=1}^{n} {\rho^2_k \over (x-z_k)^2}
  \right] ,
  \label{f}
\end{eqnarray}
where $z_i$ is the position and $\rho_i$ is the
size of the $i$-th instanton ($k=1, \cdots, n$), and all the
instantons have the same color orientations. For the $n$-instanton,
$f=\ln
\phi$ and
$\phi$ are singular at $n$ points, 
$x=z_k (k=1, \cdots, n)$.  The gauge potential reads
\begin{eqnarray}
  {\cal A}_\mu^A(x) = 
  {-2\sum_{k=1}^{n} {\rho^2_k \over (x-z_k)^4} \eta_{\mu\nu}^A 
  (x-z_k)_\nu \over 1+ \sum_{k=1}^{n} {\rho^2_k \over (x-z_k)^2}} .
  \label{singularsol}
\end{eqnarray}
The one-instanton solution \cite{BPST75} is obtained as a special
case, $n=1$, 
\begin{eqnarray}
  {\cal A}_\mu^A(x) = 
  {-2 \rho^2 \eta_{\mu\nu}^A 
  (x-z)_\nu \over  (x-z)^2[(x-z)^2 + \rho^2]} .
  \label{1singularsol}
\end{eqnarray}
The singular solution behaves as a pure gauge near the singular point
$x=z$, 
\begin{eqnarray}
  {\cal A}_\mu(x):= {\cal A}_\mu^A(x) {\sigma^A \over 2} 
  \rightarrow U^\dagger(y) \partial_\mu U(y)   ,
\end{eqnarray}
since ${\cal A}_\mu$ is written as
\begin{eqnarray}
  {\cal A}_\mu(x) 
  = U^\dagger(y) \partial_\mu U(y) {\rho^2 \over y^2+\rho^2} ,
\end{eqnarray}
with $y=x-z$, 
\begin{eqnarray}
   U(x) = {x_4 + i x_A \sigma^A \over |x|} ,
   \quad
   |x| := \sqrt{x_4 x_4 + x^A x^A} .
\end{eqnarray}
Since the self-duality and field equations are
gauge covariant, the gauge transformed potential also satisfies them.
This holds also for the monopole equation (\ref{MAg}).
By the appropriate inverse gauge transformation $U$,
we can get rid of the singularity and the resulting solution vanishes
at $x=z$.
In fact, the singularity at $x=z$ can be removed by a singular gauge
transformation,
\begin{eqnarray}
  {\cal A}_\mu{}'(x)
  = U(y)[{\cal A}_\mu(x)   +\partial_\mu]U^\dagger(y) ,
\end{eqnarray}
and the non-singular solution is obtained \cite{Rajaraman89}
\begin{eqnarray}
  {\cal A}_\mu^A{}'(x)
  =   {-2  \bar \eta_{\mu\nu}^A (x-z)_\nu \over  (x-z)^2 +\rho^2} ,
  \label{1regularsol}
\end{eqnarray}
where $\bar \eta$-symbol is defined by
\begin{eqnarray}
   \bar \eta_{\mu\nu}^A := \epsilon_{A\mu\nu}
  - \delta_{A\mu} \delta_{4\nu} + \delta_{A\nu}\delta_{4\mu} .
\end{eqnarray}
Indeed, the solution (\ref{1regularsol}) has no singularity at any
$x$. The non-singular solution approaches the pure gauge as $x
\rightarrow
\infty$,
\begin{eqnarray}
  {\cal A}_\mu{}'(x)   \rightarrow U(x) \partial_\mu U^\dagger(x)  .
\end{eqnarray}
Note that the multi-antiinstanton is obtained by
interchanging 
$\eta$-symbol and
$\bar \eta$-symbol which is self-antidual,
\begin{eqnarray}
 \bar \eta_{\mu\nu}^A  = - {1 \over 2} \epsilon_{\mu\nu\alpha\beta}
   \bar \eta_{\alpha\beta}^A .
\end{eqnarray}
\par
If we restrict the diagonal component
\begin{eqnarray}
  {\cal A}_\mu^3(x) = \eta_{\mu\nu}^3 \partial_\nu f(x) 
  =  \epsilon_{3\mu\nu} \partial_\nu f(x)  ,
\end{eqnarray}
and $f$ is independent from $4$, 
\begin{eqnarray}
  {\cal A}_i^3(x) := a_i(x) =  \epsilon_{ij} \partial_j f(x)  
  (i,j=1,2) ,
\end{eqnarray}
This solution is similar to the Witten solution \cite{Witten77} for
the multi-instanton with cylindrical symmetry, see section V of
\cite{KondoII}.
\par
It is known \cite{MW96} that various well-known equations in lower
dimensions are obtained by (dimensional) reduction of the self-dual
equation in four dimensions.  For example, if the field is static,
i.e., 
$\partial_0 {\cal A}_\mu=0$, the identification 
${\cal A}_0(x_0,{\bf x}) = \phi({\bf x})$ with 
${\bf x}=(x_1,x_2,x_3)$ leads to the Bogomol'nyi equation describing
the magnetic monopole in three dimensions,
\begin{eqnarray}
  F_{ij}({\bf x}) = \pm \epsilon_{ijk} D_k \phi({\bf x})  \quad
(i,j,k=1,2,3) .
\end{eqnarray}
In this sense, the self-dual equation contains a kind of magnetic
monopole.

\par
 Under the ansatz (\ref{ansatz}), the self-dual equation (\ref{SDYM})
for the instanton is satisfied only for the function $f$ given by
(\ref{f}) which is also a solution of the Yang-Mills field equation 
(\ref{Feq}). The same ansatz gives a solution of the monopole
equation for arbitrary
$f$ which is not necessarily the solution of the Yang-Mills field
equation.
The general solution of the self-dual equation is given according to
the method of Atiyah, Drinfeld, Hitchin and Mannin (ADHM). 
To author's knowledge, the general solution is not known for the
monopole equation (\ref{MAg}). 
In order for the solution of (\ref{MAg}) to give a magnetic
monopole, we must check whether the solution gives non-trivial
$k_\mu$.  The explicit monopole solution was constructed in 
 \cite{CG95} and \cite{BOT96}.  
Only the solution of Brower, Orginos and Tan (BOT) 
\cite{BOT96} satisfies ${\cal R}[{\cal A}]<\infty$.
For more details, see \cite{BOT96}.

\par
\section{Comparison with the Cho-Faddeev-Niemi variables}

\par
If we choose 
\begin{eqnarray}
{\cal R}[{\cal A}] := \int d^Dx \ {\rm tr}_{G}
 \left( {1 \over 2}{\cal A}_\mu(x) {\cal A}_\mu(x) \right) ,
\end{eqnarray}
the variation is given by
\begin{eqnarray}
 \delta_\omega {\cal R}[{\cal A}] := \int d^Dx \ {\rm tr}_{G}
 \left( {\cal A}_\mu(x) \delta_\omega {\cal A}_\mu(x) \right) 
 = \int d^Dx \ {\rm tr}_{G}
 \left( {\cal A}_\mu(x) D_\mu[{\cal A}]\omega(x) \right) 
 \\
 = - \int d^Dx \ {\rm tr}_{G}
 \left( D_\mu[{\cal A}]{\cal A}_\mu(x) \cdot \omega(x) \right) ,
\end{eqnarray}
where we have used the partial integration by parts.
The requirement $\delta_\omega {\cal R}[{\cal A}]=0$ for
arbitrary $\omega$ yields  
$0=D_\mu[{\cal A}]{\cal A}_\mu(x)= \partial_\mu {\cal A}_\mu(x)$.
This is the familiar Lorentz gauge which is a linear gauge.  The
linear equation $\partial_\mu {\cal A}_\mu(x)=0$ can not have a
soliton solution. 
\par
Another way to obtain $\delta_\omega {\cal R}[{\cal A}]=0$ is to
restrict the gauge transformation $\omega$ such that 
$D_\mu[{\cal A}]\omega(x)=0$.  This equation is solved for 
${\cal A}_\mu$, see \cite{Cho80} .
For example, in the case of $G=SU(2)$,
\begin{eqnarray}
 D_\mu[{\cal A}]\omega(x) := \partial_\mu \omega^A(x)
 + g \epsilon^{ABC} {\cal A}_\mu^C(x) \omega^B(x) = 0 .
\end{eqnarray}
A solution is given by
\begin{eqnarray}
 {\cal A}_\mu^A(x) = a_\mu \omega^A(x) - {1 \over 2g} 
 \epsilon^{ABC} \omega^B(x) \partial_\mu \omega^C(x) ,
\end{eqnarray}
where $a_\mu$ is arbitrary Abelian vector field and $\omega^A$ is
chosen to be a unit vector in three dimensions,
$\omega^A(x) \omega^A(x) = 1$.
In this way, it is possible to obtain a
subset of a non-Abelian gauge theory. This formalism gives  
essentially the same result as the partial gauge fixing, the MA
gauge.   The details will be given in a forthcoming publication.

\section*{Acknowledgments}
The author would like to thank the referee for giving constructive comments in revising the paper.
This work is supported in part by
the Grant-in-Aid for Scientific Research from the Ministry of
Education, Science and Culture (No.10640249).


\end{document}